\pdfoutput=1
\documentclass{article}
\usepackage[preprint]{neurips_2026}

\usepackage[utf8]{inputenc} 
\usepackage[T1]{fontenc}    
\usepackage{hyperref}       
\hypersetup{
  hidelinks,
  pdftitle={The Jagged Global Economy: Frontier AI Unevenly Exposes National Economies},
  pdfauthor={Arul Murugan, Tomás Aguirre, Abhishek Nagaraj, Rishi Bommasani},
  pdfsubject={National AI exposure and cross-country labor-market heterogeneity}
}
\usepackage{url}            
\usepackage{booktabs}       
\usepackage{amsfonts}       
\usepackage{nicefrac}       
\usepackage{microtype}      
\usepackage{xcolor}         
\usepackage{graphicx}
\usepackage{array}
\usepackage{float}
\usepackage{placeins}

\usepackage{amsmath,amsfonts,bm}

\newcommand{\captiona}{{\em (a)}}
\newcommand{\captionb}{{\em (b)}}
\newcommand{\captionc}{{\em (c)}}

\def\eqref#1{equation~\ref{#1}}

\def\1{\bm{1}}

\DeclareMathAlphabet{\mathsfit}{\encodingdefault}{\sfdefault}{m}{sl}
\SetMathAlphabet{\mathsfit}{bold}{\encodingdefault}{\sfdefault}{bx}{n}

\usepackage{pdflscape}
\usepackage{pifont}
\usepackage{xspace}
\usepackage[normalem]{ulem} 

\newcommand\tldrDone[1]{}

\newcommand\eg{e.g.\xspace}
\newcommand\ie{i.e.\xspace}

\def\Snospace~{\S{}}

\title{The Jagged Global Economy:\\
Frontier AI Unevenly Exposes National Economies}

\author{
Arul Murugan\\
UC Berkeley\\
\And
Tomás Aguirre \\
University of São Paulo \\
\And
Abhishek Nagaraj\\
UC Berkeley\\
\And
Rishi Bommasani\\
Stanford University\\
}

\begin{document}

\maketitle
\begin{abstract}
Frontier AI's labor-market effects matter to workers, firms, and policymakers, but current evidence generally comes from a handful of high-income economies.
The capabilities of frontier AI are jagged across work tasks and national economies diverge in how they allocate human labor.
We introduce a national AI exposure metric that combines occupation-level exposure scores and international employment data for 141 countries. We find that high income countries are substantially more exposed than low income countries and that Europe \& Central Asia are 50\% more exposed than Sub-Saharan Africa.
We also find a gender gap: women are more exposed than men in 91\% of countries, driven by their concentration in white-collar and sales occupations. The exceptions are countries where women's employment remains concentrated in agriculture and household enterprises.
We validate our national AI exposure estimates by showing they predict national AI adoption statistics published by Anthropic, Microsoft, and OpenAI.
Beyond direct exposure, we identify a new mechanism for indirect exposure due to cross-country income dependencies.
Some nations such as Tajikistan depend heavily on foreign workers remitting money back to their home countries: Tajikistan's direct exposure to frontier AI is below-average but because 37\% of Tajikistan GDP is Russian remittance and Russia is very exposed, Tajikistan's remittance-accounted exposure becomes above-average.
Our research shows that national variation in exposure is large enough that policy responses calibrated to U.S. or European labor markets will not generalize.
\end{abstract}

\section{Introduction}
How frontier AI will reshape national economies is central to workers, firms, and governments.
Around the world, the public already expects AI to reshape work: a majority of respondents in the 2025 Stanford AI Index report that AI will change how people do their jobs within five years, while the 2025 Ipsos AI Monitor finds that more people think AI will worsen than improve their local job market \citep{hai2025aiindex,ipsos2025aimonitor}. 
Policymakers frame frontier AI as a strategic priority \citep{uk2025aiopportunities,canada2025sovereigncompute,ec2025aicontinent} and, reciprocally, frontier AI firms target national AI sovereignty agendas \citep{openai2025countries}.
AI investment is seen as critical for future growth: the UN projects that the global AI market will expand to \$4.8 trillion by 2033 \citep{unctad2025tir}.

These effects are likely to be very unevenly distributed across countries. 
Frontier AI capabilities are jagged across tasks \citep{dellacqua2026jagged}, and national economies differ substantially in how they allocate workers to jobs.
Consider call center work. \citet{brynjolfsson2025generative} find that deploying generative AI in a large call center increases issues resolved per hour by 15\% with larger gains for less experienced and lower-skilled workers.
A productivity shock of this level at scale would disproportionately disrupt nations like India or the Philippines where a large share of the national labor force is tied to call center work.
The same frontier AI technologies could complement workers in one nation, reorganize entry-level labor markets in another, and have little effect in a third.

\begin{figure}[t]
    \centering
    \includegraphics[width=\linewidth]{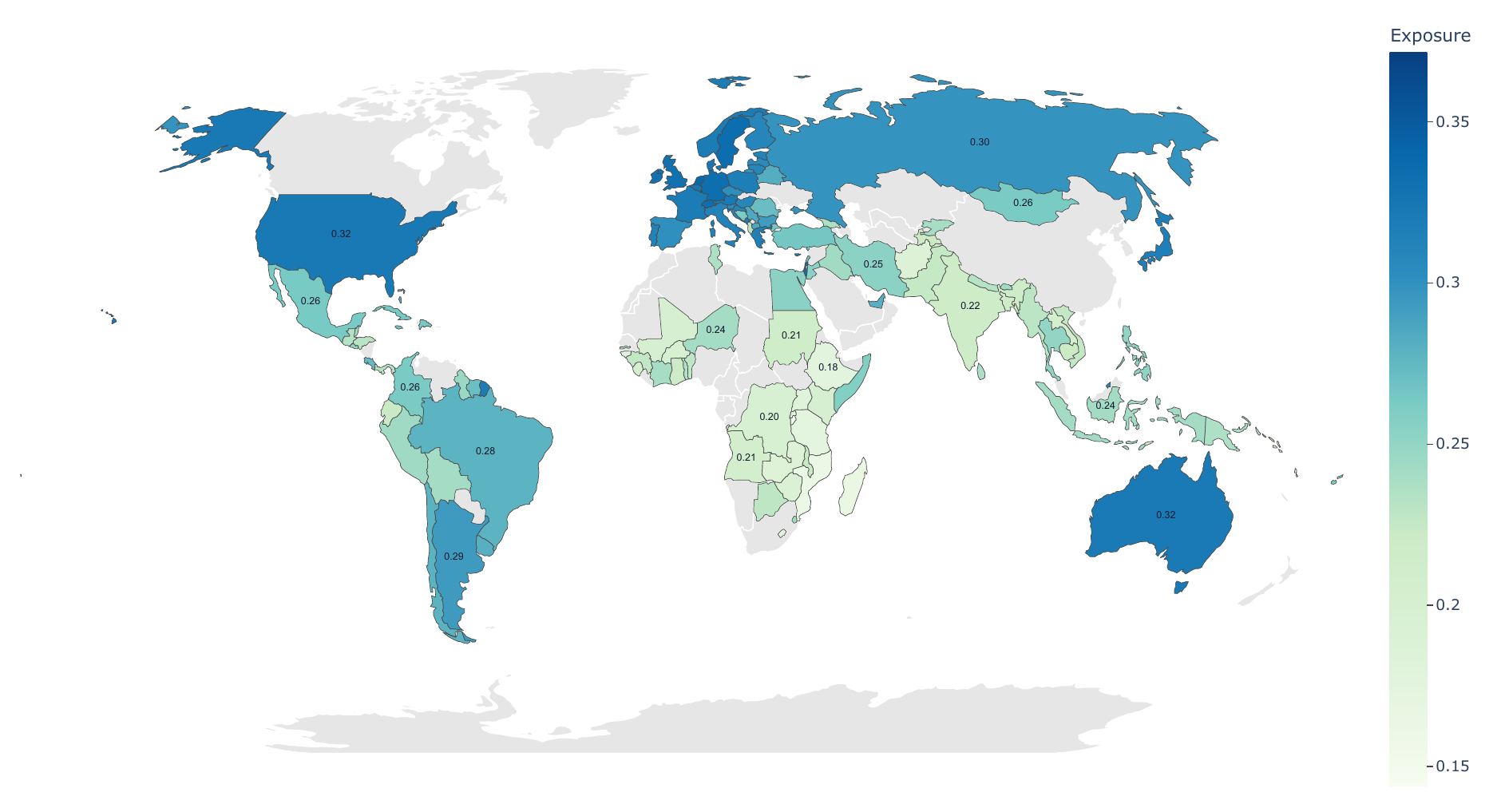}
    \caption{\textbf{National AI exposure varies substantially across countries.}
    Darker countries are more exposed to frontier AI. Countries without 2-digit ISCO-08 employment data are shown in gray.}
    \label{fig:national_exposure_distribution}
\end{figure}

We introduce a national AI exposure metric to capture this heterogeneity. 
Our approach combines occupational exposure estimates from the economics of frontier AI literature with internationally comparable statistics on how countries allocate human labor across occupations.\footnote{Concurrent with our work, \citet{gmyrek2026disruption} introduce a national AI exposure metric with a similar methodology but do not publish per-country exposure estimates at the time of writing.} 
The metric summarizes how strongly a national economy’s current labor allocation aligns with the tasks that frontier AI can already accelerate or transform. 
The resulting object is not a forecast of adoption, wages, or net employment losses. 
Rather, it is a comparable measure of exposure at the national level: a way to quantify which countries are more exposed, through the composition of their labor markets, to the jagged capabilities of frontier AI.
This exposure is not necessarily positive or negative: it signals opportunities for productivity gains but these gains are not certain, they may contribute to economic growth, and they may reduce the comparative advantage of human labor but could also increase it via certain patterns of augmentation.

We find substantial variation in national exposure (\autoref{fig:national_exposure_distribution}): the most exposed nation (Luxembourg) is $2.6\times$ more exposed than the least exposed nation (Burundi).
The most exposed regions are North America and Europe \& Central Asia, which are at least 50\% more exposed than the least-exposed region of Sub-Saharan Africa.
We study different explanations for this result and find that disparities in the prevalence of white collar work explain much of the observed variation.

Given the overall trends in national exposure, we disaggregate to study specific country and worker characteristics.
At the country level, we find that higher income countries are more exposed.
At the worker level, we find a pervasive gender gap: women in 91\% of studied nations are more exposed than men because of their occupational composition.
In many countries, women already face barriers to labor force entry and receive lower wages for comparable work product \citep{ilo2024careparticipation, blau2017genderwagegap}.
The groups we find to be more exposed (\ie richer countries, women) also poll less optimistically about AI work \citep{mcclain2025publicexpertsai}, so our results may explain why these groups are less optimistic.

We test the predictive validity of our national AI exposure metric by predicting national AI adoption according to usage statistics published by Anthropic, Microsoft, and OpenAI.
We find strong monotonic relationships between our exposure estimates and all of these adoption statistics.
We estimate that $0.10$ increase in national AI exposure corresponds to $12\times$ growth in national per-capita Claude usage and 19 percentage-point increase in the national generative AI adoption rate.

Since the economic impacts of frontier AI are not only determined by domestic activity but also cross-country relationships, we study indirect exposure channels. 
We identify a novel mechanism for national exposure: foreign workers sending remittance back to their families in their home countries.
We show that this channel substantially increases national exposure for several countries where foreign remittance is a sizable fraction of national GDP.
For example, 25\% of the GDP of Honduras, Guatemala, and El Salvador is remittance and more than 80\% for all three countries is remitted from the United States.
Since the national AI exposure of these three Central American countries is below average but the United States is highly exposed, accounting for this indirect effect substantially raises national AI exposure in Honduras, Guatemala, and El Salvador.
National policymakers in these countries should track how the foreign labor economies they depend on are changing because of frontier AI adoption and the resulting second-order effects on their national income.
Overall, our research identifies critical heterogeneity that should inform national AI and economic policy.
\section{Data}
To estimate national exposure to frontier AI, we combine occupational employment statistics from the International Labour Organization (ILO) with occupational exposure estimates from Gmyrek et al.

\subsection{Occupational employment statistics}
To study the impact of AI on labor economies around the world, we use country-level data compiled by the ILO.
The ILO publishes employment statistics based on the International Standard Classification of Occupations (ISCO-08), which is a four-level hierarchy that spans 436 occupational categories.
As an example, one of the 436 occupational categories is ``Generalist Medical Practitioners'' (code 2211; code 2 = Professionals, code 22 = Health Professionals, code 221 = Medical doctors).\footnote{For brevity, we will sometimes refer to occupational categories as ``occupations'' where the distinction is immaterial.}

We use the most recent annual statistics for each country's employment counts for the 43 2-digit (sub-major group) ISCO-08 occupational categories, along with available ILO wage statistics for wage-weighted extensions.
The counts are produced by national statistical offices through country-level data collection (\eg labor force surveys, census, administrative records) and then compiled and disseminated by the ILO.
We describe data processing in \autoref{app:data}, including the country filters and the exclusion of armed-forces, aggregate, and residual occupational rows.
We end up with employment data for 40 occupational categories across 141 countries.
This provides strong coverage of the world's nations, with most missing nations being very small countries with less mature government labor data infrastructure.\footnote{We lack 2-digit ISCO-08 data for a few major economies (\eg China, Canada, Saudi Arabia).}


\begin{figure}[ht]
    \centering
    \includegraphics[width=0.75\linewidth]{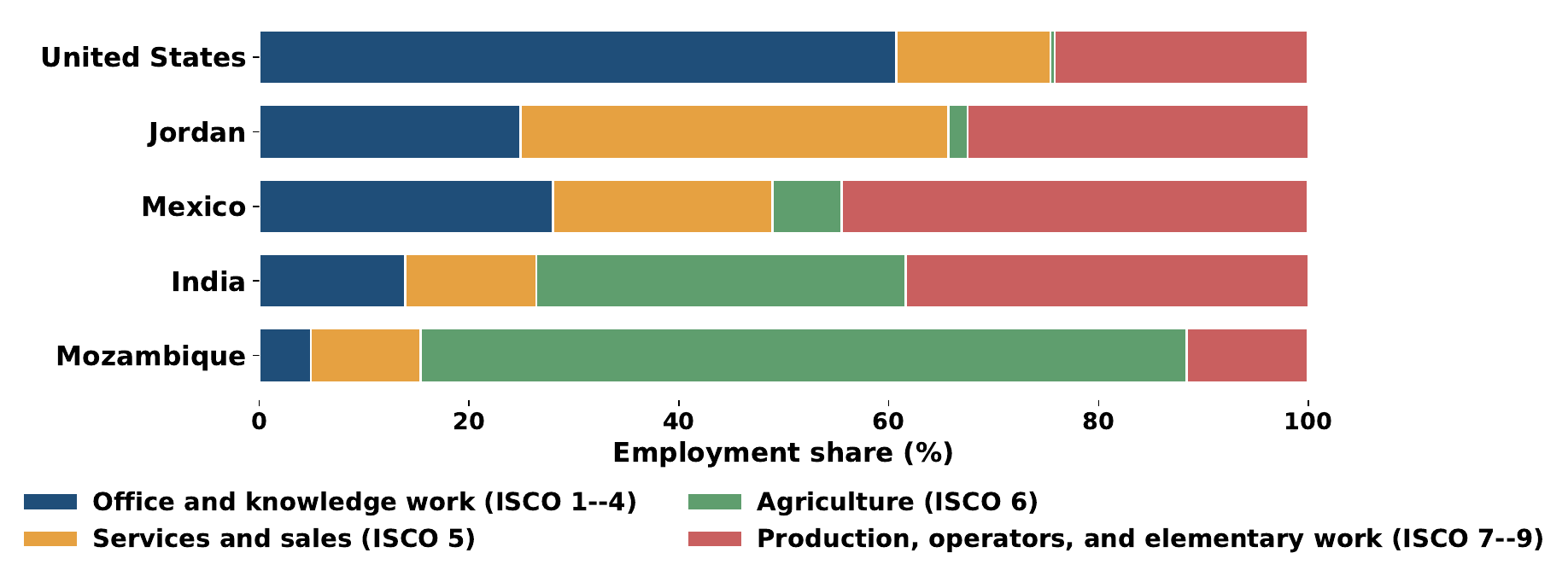}
    \caption{\textbf{Illustrative cross-country differences in labor composition.}}
    \label{fig:cross_country_labor_composition}
\end{figure}

\autoref{fig:cross_country_labor_composition} visualizes how employment composition varies around the world.
For example, office- and knowledge-intensive work (ISCO major groups 1--4) accounts for 61\% of employment in the United States, but only 14\% in India and 5\% in Mozambique. 
And agriculture (ISCO 6) accounts for 73.0\% of employment in Mozambique and 35\% in India but only 0.4\% in the United States. 
Beyond labor compositional differences, wages also vary greatly: the average monthly white-collar wage is PPP \$7{,}778 in the United States, compared with PPP \$1{,}740 in India.\footnote{PPP denotes purchasing-power-parity-adjusted U.S. dollars to account for local price levels across countries.}
White-collar work earns 74.2\% of the wage bill in the United States to 30.4\% in India.

\subsection{Occupational exposure estimates}
AI capabilities can impact labor through multiple mechanisms.
A critical mechanism is \textit{productivity}. 
AI could reduce the time spent performing tasks.
\textit{Exposure} measures the time spent by a worker performing a task (or occupation) that \textit{could} be saved by current technological capabilities.
Exposure estimates bridge technological progress with economic task-based models, a standard framework in the economics of technological change that studies how technologies substitute for, complement, and create tasks rather than affecting occupations as undifferentiated bundles \citep{autor2003skill,acemoglu2011skills,acemoglu2019automation}.

\citet{eloundou2024gpts} popularized the application of exposure to the economic impacts of frontier AI. 
To estimate occupational exposure, they average across an occupation's tasks based on whether AI capabilities would save half the time spent on the task while preserving quality.
To encode the level of capability at the time, they specified primitives that models could do (\eg summarize medium-length documents) and could not do (\eg repair physical equipment).
Their empirical estimates use human and GPT-4 annotations to determine task exposure, leading to an overall estimate that ``80\% of the U.S. workforce could have at least 10\% of their work tasks affected by the introduction of LLMs, while approximately 19\% of workers may see at least 50\% of their tasks impacted.''
Prior to the work of \citet{eloundou2024gpts}, earlier works estimated occupational exposure to machine learning and AI based on occupational abilities \citep{felten2018method,felten2021occupational}, patents \citep{webb2020impact}, and by mapping cognitive abilities required for tasks to benchmarks \citep{tolan2021measuring}.
Following their work, others have expanded exposure estimates to multimodal capabilities \citep{felten2023occupational}, complementarity rather than pure substitution \citep{pizzinelli2023labor}, and to integrate evidence from observed usage \citep{massenkoffmccrory2026labor}.

While exposure pervades empirical research on frontier AI, critiques target (i) the aggregation from tasks to occupations or (ii) the divergence between theorized and observed exposure.
The standard aggregation of averaging task-level exposure to yield occupation-level exposure may incorrectly imply a linear production structure: occupations with the same average exposure may face different impacts when exposed tasks appear in chains or when tasks are quality complements and bottlenecks \citep{demirer2026chaining,gans2026oring}.
Further, the impact on wages or employment may be better predicted not by the tasks that are automated, but by the expertise requirements of the residual tasks \citep{autor2025expertise}.
\citet{eloundou2024gpts} emphasize that exposure quantifies technical feasibility rather than observed productivity: \citet{narayanan2025ainormaltechnology} posit a variety of diffusion bottlenecks and \citet{massenkoffmccrory2026labor} quantify large gaps between theoretical exposure and observed usage in several occupations.

Most AI exposure estimates use the U.S. occupational classification system.
Our focus is global and not limited to the U.S., so we use exposure estimates for the international ISCO occupational classification system prepared by researchers at the ILO \citep{gmyrek2025exposure}. 
Since the ISCO occupations do not have rich task-level decompositions, they first use Polish occupational data that maps occupations to 29,753 distinct tasks and gather human judgments about the automation potential of a representative sample of 2,861 tasks from 1,640 workers. 
Experts review these worker assessments and they use the resulting expert-adjusted annotations to train a model to predict task-level exposure, which is then aggregated for all 4-digit ISCO-08 occupations.
\section{National AI Exposure Estimates}
We quantify national exposure to frontier AI based on nation-agnostic occupational exposure estimates and nation-specific occupational employment statistics.\footnote{This neglects heterogeneity (\eg task composition) in the same occupation across countries. However, intra-occupation heterogeneity may be dominated by intra-category cross-occupational heterogeneity at the 2-digit ISCO level. \citet{gmyrek2026disruption} attempt to address this via a more complex nation-sensitive estimation procedure for occupation exposure.}
Prior work applies similar approaches using different estimates for occupational AI exposure \citep{cazzaniga2024genai, lewandowski2025workers, gmyrek2026disruption}.
We define national AI exposure as the employment-share-weighted average exposure of a country's occupational structure.

\noindent For an occupation $j$, let $e_j$ quantify its exposure and $c_{i, j}$ count its employment in nation $i$.
\begin{align}
    {n}_i &= \frac{\sum_j c_{ij} e_j}{\sum\limits_j c_{ij}}
\end{align}
National AI exposure varies substantially across countries (\autoref{fig:national_exposure_distribution}): the most-exposed country (Luxembourg, ${n} = 0.37$) is $2.6\times$ more exposed than the least-exposed country (Burundi; ${n} = 0.14$).
Luxembourg's unusually large finance-and-insurance sector drives its high exposure: business and administration professionals and associate professionals together account for 38.7\% of total exposure, with legal, social and cultural professionals contributing another 10.0\%. This pattern is consistent with Luxembourg's unusually large finance-and-insurance sector.\footnote{Luxembourg had the European Union's highest employment share in finance and insurance \citep{eurostat2024financeinsurance}.} 
Singapore also demonstrates high exposure through a similar but more technical/managerial mechanism: national exposure is 0.37, the white-collar share is 73.8\%, and business, administration, and managerial occupations together account for 45.1\% of total exposure \citep{mom2026labourforcesingapore}.
At the bottom, Burundi shows how strongly the index is pulled down by agrarian employment structure. 
It has the sample's lowest exposure (0.1440), the lowest white-collar share (3.1\%), and a strongly negative white-collar residual (-0.0468). 
Nearly 69.3\% of Burundi's total exposure comes from subsistence farmers, fishers, hunters and gatherers alone \citep{worldbank2024burundiyouth}.
\autoref{app:alternatives} shows that the results are robust to the choice of occupational exposure estimates.

On top of the large variation in national exposure, we find a persistent \textit{gender gap}: women are more exposed than men in 126 of the 138 countries we study.
The median gender gap is 13.7\% of national exposure and is driven by occupational composition.\footnote{\citet{manning2026adaptable} observe this pattern for the U.S. labor economy.}
Cross-country differences in male and female white-collar shares explain about half of the variation in the gender gap (50.3\%), and including sales raises that explanatory power to 87.8\%, indicating that women's greater concentration in white-collar and sales work is the dominant structural mechanism.
The pattern is visible in concrete cases. 
Women are substantially more exposed in the Philippines (0.2964 for women versus 0.2214 for men) and Jamaica (0.3173 versus 0.2276), where official statistics show women concentrated in service-and-sales, clerical, and professional work \citep{psa2024womenmenwork, statin2020jamaicanlabourmarket}.
By contrast, women are less exposed in Pakistan (0.1970 for women versus 0.2338 for men) and India (0.2001 versus 0.2240), where women's employment remains more concentrated in agriculture and household-enterprise work \citep{pbs2026lfs202425, mospi2023womenmenindia}.
This result that women may be more prone to AI-mediated labor disruption warrants concern given preexisting challenges for women to find gainful employment.
In many countries, women experience barriers to labor force participation \citep{ilo2024careparticipation} and receive lower wages than men, including within occupations and after accounting for standard observable characteristics \citep{blau2017genderwagegap}.
And on the topic of AI, women are less likely to adopt the technology for work \citep{humlum2025unequal,otis2024globalgendergapsgenai} and have less positive impressions of the technology \citep{mcclain2025publicexpertsai}.
Consequently, women workers may be more disrupted by AI because of men adopting AI in work.
\subsection{Explaining national AI exposure}
\label{sec:predicting-exposure}
\begin{figure}[htbp]
    \centering
    \includegraphics[width=\linewidth]{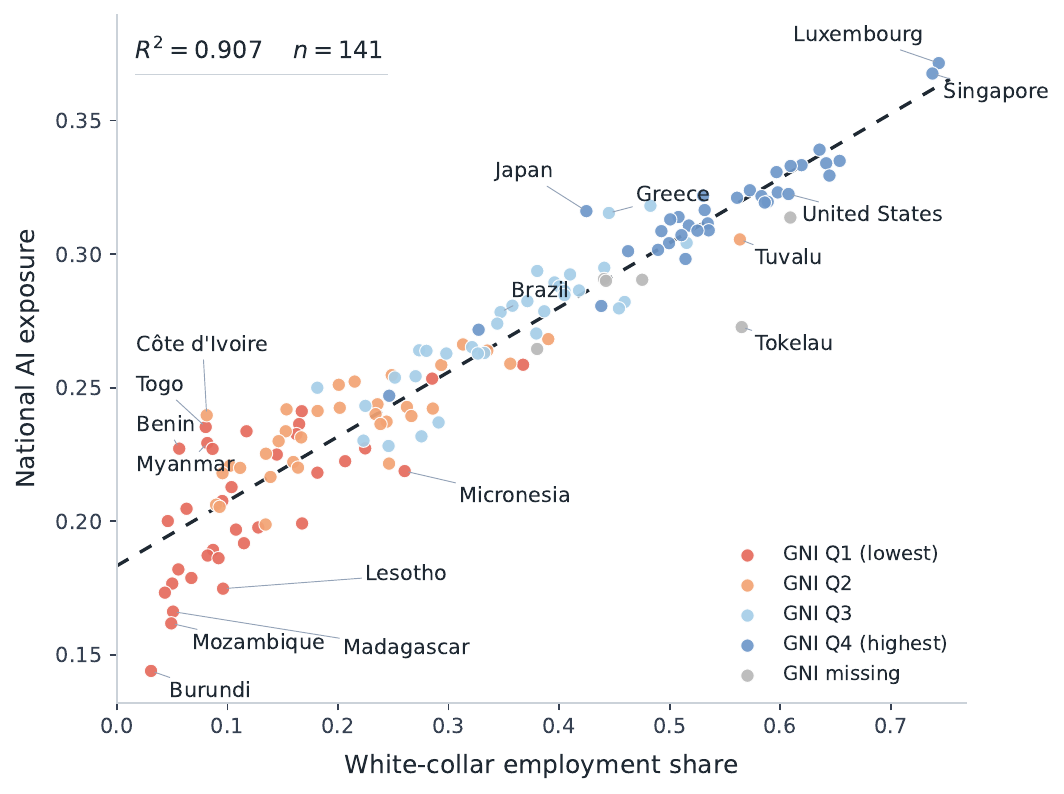}
    \caption{\textbf{White-collar employment share predicts national AI exposure.}}
    \label{fig:wc_share_vs_exposure}
\end{figure}

Cross-national variation in labor composition yields substantial differences in national AI exposure.
Can these differences be explained by more basic properties of the countries or their workforces?
\autoref{fig:wc_share_vs_exposure} shows national exposure as a function of the white collar share of the country's labor force.
We find that the white collar share is highly predictive of exposure ($R^2 = 0.91$).
We also find that gross income is predictive of exposure, but less so than white collar share: log gross national income (log GNI) explains 77\% of cross-country variation in national AI exposure.
Together, the two predictors have an $R^2$ of $0.93$, indicating that the workforce-specific white collar share statistic is more explanatory than the broader national-level income statistic.
We further explore how the exact classification of occupations as white collar or blue collar affects this result: changing the label of a single occupational category by classifying sales occupations (ISCO-08 code 52) as white collar instead of blue collar substantially increases explanatory power ($R^2 = 0.96$).
In \autoref{app:predictivity}, we test other predictors, generally finding that measures related to computer use, written comprehension, administrative work, and information processing are good predictors.

The correlation between national AI exposure, white collar work, and national income manifests via other channels.
Cross-country surveys find substantial heterogeneity in public opinion on AI on matters spanning trust in the technology, confidence in its benefits, concerns about its risks, and desire for regulation.
For example, respondents in advanced economies generally report less trust and perceive less benefit from frontier AI relative to respondents from developing economies \citep{pew2025globalai,kpmg2025aiglobal,ipsos2025aimonitor}.
Anthropic interviewed 81k Claude users, finding that workers in occupations with more observed exposure are more likely to express concern about job displacement \citep{massenkoffhuang2026economics}.
Beyond heterogeneity, many of these studies do report an overarching concern that the public fears and expects AI-driven job loss \citep{ipsos2025predictions}, which reflects a common interpretation of exposure as displacement risk even if exposure need not imply worse labor-market outcomes \citep{imas2026howwillai}.

\section{Predicting National AI Adoption}
While exposure estimates translate from technological capabilities to economic quantities, they can be prone to misunderstanding \citep{eloundou2024gpts, manning2026adaptable, imas2026howwillai}.
Exposure identifies where AI could plausibly increase productivity, not which jobs will necessarily disappear.
For past technologies, economics research substantiates that task- and occupation-level exposure estimates predict downstream labor outcomes \citep{webb2020impact,autor2025expertise}.
For frontier AI, emerging evidence also shows that exposure estimates predicted observed frontier AI adoption \citep{tomlinson2025workingwithai,chatterji2025howpeopleusechatgpt,handa2025economictasks} and AI-mediated employment reduction \citep{brynjolfsson2025canaries,  tucker2026qwi}.

We test whether national AI exposure estimates predict observed national outcomes.
As \citet{chandar2025ailabormarkets} notes, current data is insufficient to study broad cross-national wage or employment changes attributable to AI.
Therefore, we focus our attention on predicting national AI adoption using statistics published by frontier AI companies based on chatbot usage data.
We use data from Anthropic's Anthropic Economic Index for Claude usage \citep{anthropic2026aeiv5}, Microsoft's AI Diffusion adoption measures \citep{microsoft2026aidiffusion,microsoft2026aidiffusiondata}, and OpenAI's ChatGPT usage statistics \citep{chatterji2025howpeopleusechatgpt,openai2026signalsglobal} with details in \autoref{app:data_usage}.

\begin{figure}[H]
    \centering
    \begin{minipage}[t]{0.32\linewidth}
        \centering
        \includegraphics[width=\linewidth]{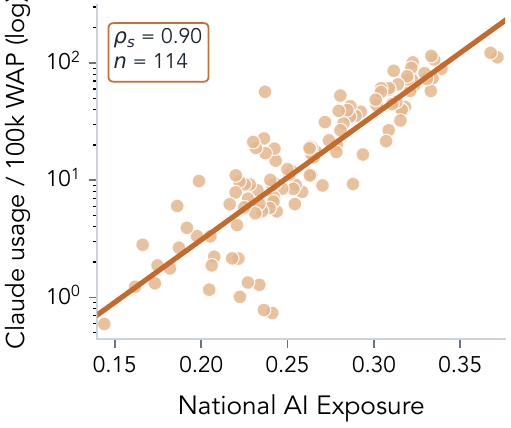}
        {\footnotesize \captiona\ Anthropic Claude usage}
    \end{minipage}\hfill
    \begin{minipage}[t]{0.32\linewidth}
        \centering
        \includegraphics[width=\linewidth]{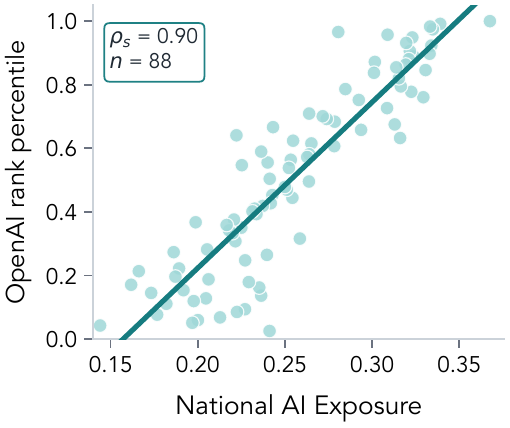}
        {\footnotesize \captionb\ OpenAI Signals rank percentile}
    \end{minipage}\hfill
    \begin{minipage}[t]{0.32\linewidth}
        \centering
        \includegraphics[width=\linewidth]{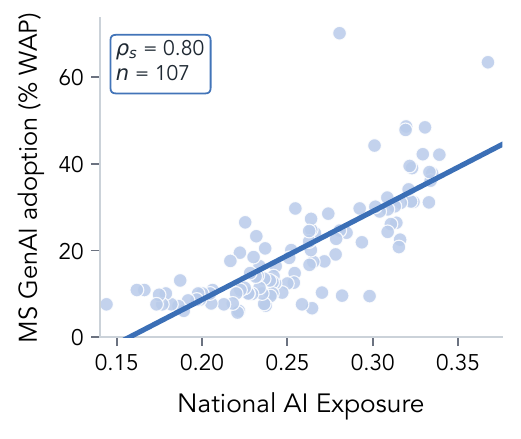}
        {\footnotesize \captionc\ Microsoft GenAI adoption}
    \end{minipage}

    \caption{\textbf{National AI exposure predicts national AI adoption.}
    \captiona\ Anthropic Claude usage per 100,000 working-age people, using the Anthropic Economic Index Claude.ai country release for February 5--12, 2026, and plotted on a log scale.
    \captionb\ OpenAI Signals country-rank percentile for calendar-year 2025, based on OpenAI's population-normalized ranking of countries by per-capita share of sampled consumer ChatGPT messages; OpenAI publishes ranks rather than country-level usage rates.
    \captionc\ Microsoft AI Diffusion adoption rate for Q1 2026, measured as the percent of the working-age population using any of the covered generative-AI tools.}
    \label{fig:observed_outcomes_vs_exposure}
\end{figure}

We find that our national AI exposure estimates predict all three observed AI usage measures, although the three companies publish different types of statistics. 
Using the Anthropic data, we find that national Claude usage scales approximately exponentially in national AI exposure: in the fitted relationship, a 0.10 increase in national AI exposure yields a $12\times$ increase in national Claude usage per 100,000 working-age people. 
Since OpenAI ranks countries rather than publishing per-country statistics, we study ranking correlation within the 88 countries shared across our data and OpenAI Signals.
The top quintile by exposure (average exposure percentile = 90\%) has high ChatGPT usage with an average OpenAI percentile of 86.5\%.
Conversely, the bottom quintile by exposure (average exposure percentile = 10\%) has low ChatGPT usage with an average OpenAI percentile of 14.9\%.
Using the Microsoft data, we find that national generative AI adoption rate\footnote{Adoption counts visits to generative AI sites/apps: Alice, ChatGPT, Character.ai, Claude, ClOVA X, DeepSeek, ERNIE Bot (Yiyan.baidu), GigaChat, Google Gemini, Grok, Khanmigo.ai, Meta.ai, Microsoft Copilot, Midjourney, Mistral.ai, NanoSemantics AI Assistant, Perplexity, Tongyi Qianwen, and Xiaowei.} scales linearly in national AI exposure: in the fitted relationship, a 0.10 increase in national AI exposure yields an additional 203 national adopters per 1,000 working-age people.

Given our previous results (\autoref{sec:predicting-exposure}) that national AI exposure is itself predictable from national white-collar share and gross national income (GNI), we test whether national AI exposure provides additional explanatory power over these standard national metrics in predicting national AI adoption.
Exposure alone strongly predicts all three adoption measures, with $R^2=0.77$ for Anthropic Claude usage, $R^2=0.81$ for OpenAI Signals, and $R^2=0.61$ for Microsoft AI Diffusion.
However, once white-collar share and log GNI are included, adding national AI exposure contributes little additional explanatory power: the full-model $R^2$ is unchanged for Anthropic and Microsoft after rounding, and increases by only 0.001 for OpenAI (see \autoref{tab:appendix_predicting_adoption_regressions}).

\section{Indirect Channels for National AI Exposure}
While the direct impact of AI on domestic workers is central to determining how national labor economies change because of AI, it is not the sole determinant.
National economic outcomes depend on the activities of other nations (\eg trade, tariffs, outsourcing).
Given concerns that AI-driven automation may reduce labor demand and domestic wages \citep{acemoglu2024simple,autor2025expertise, brynjolfsson2025canaries}, we consider how this may be mediated by the impacts of AI on foreign countries.

We observe that while the concern is often expressed as AI suppressing wages, the underlying concern is that individuals will not be able to generate the income required to sustain their lives. 
In most countries, this can be restated in terms of wages, because individual income and individual wages are essentially equal.
However, we observe that in some nations, a second mechanism meaningfully contributes to income: \textit{remittance}.
Remittance is the transfer of money by a foreign worker (usually a migrant) to their home country to serve as income for their families.
We hypothesize that remittance introduces a second channel for national exposure: if income in country A is sourced via remittance from country B, then national exposure in country B indirectly exposes country A. 

To study this hypothesis, we use the KNOMAD database from the World Bank on global migration and development (\autoref{app:data_knomad}; \citealp{worldbank_knomad_bre,worldbank_knomad_mri}).
The KNOMAD country-country remittance matrix estimates annual remittance flows between source and target countries by partitioning total target-side remittance inflow proportional to bilateral migration distributions \citep{ratha2022bilateralmatrix}.
For example, since Mexico received \$52 billion from the U.S. in remittance during 2021, a decline in U.S. labor demand could indirectly reduce income in Mexico via remittance reduction.

\begin{figure}
    \centering
    \includegraphics[width=0.8\linewidth]{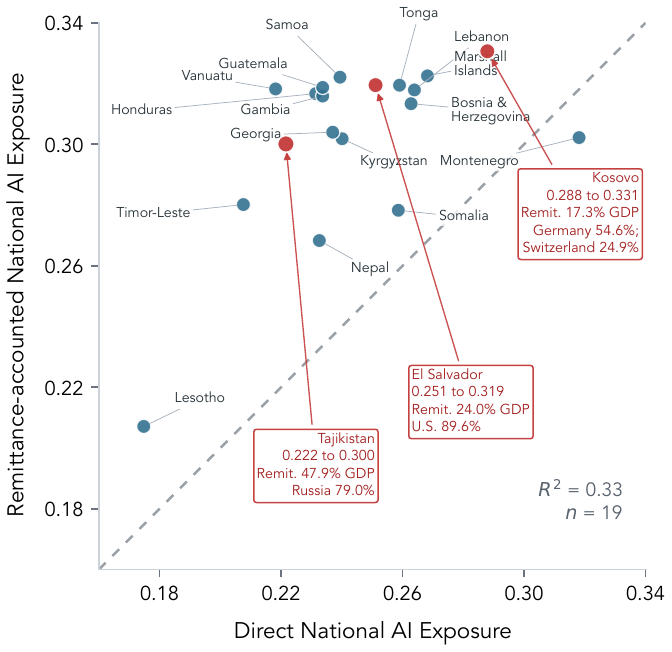}
    \caption{\textbf{Remittance-accounted vs. direct national AI exposure.}
    Countries shown have remittance of at least 10\% of national GDP.}
    \label{fig:remittance}
\end{figure}

We find that the remittance channel generally raises measured exposure among highly remittance-dependent countries: 18 of the 19 countries with remittances above 10\% of GDP move above the diagonal in \autoref{fig:remittance}.
Tajikistan is the clearest example.
It received \$6.80 billion in personal remittances in 2024, equal to 47.9\% of GDP.
The latest KNOMAD bilateral matrix is for 2021; in that matrix, Russia accounts for \$2.31 billion of Tajikistan's \$2.92 billion in measured bilateral remittance inflows, or 79.0\%.
Tajikistan's direct national AI exposure is low (${n}=0.222$, 21st percentile among countries with measured exposure), but its remittance inflows come primarily from more exposed economies, led by Russia (${n}=0.298$, 75th percentile) and Germany (${n}=0.333$, 96th percentile).
As a result, Tajikistan's remittance-accounted exposure rises to ${n}=0.300$, comparable to a 75th-percentile country in the direct exposure distribution.

This result illustrates that even if countries are not directly exposed to frontier AI, or that theoretical exposure is unlikely to convert into observed productivity (\eg due to the digital divide and lack of internet access), they should prepare for the ripple effects of AI adoption elsewhere.
In particular, while policymakers in these countries currently account for remittance in strategic planning given its scale, they may not account for the second-order remittance-mediated effect of AI on their economy.
\section{Discussion}
\label{sec:discussion}
Frontier AI development is concentrated in a few nations like the United States and China, but its economic impacts are globally diffuse.
The national AI exposure metric we introduce provides an initial lens for clarifying how frontier AI capabilities may impact countries around the world in different ways and the specific occupational categories within those labor markets.
This encodes two forms of heterogeneity: occupations are differentially exposed and national labor markets are differentially composed of occupations.
To make the most of these metrics, we believe they should be integrated with measures that operate at other levels of abstraction in reasoning about the impacts of (frontier) AI on the economy.
In turn, this introduces a third form of heterogeneity: data availability varies substantially across the world.

\paragraph{Heterogeneity in data availability.}
As we also consider in this work, usage data is especially useful in the study of frontier AI's economic impacts because it directly reflects the actual tasks and workflows where AI is adopted \citep{manning2025DOL}.
From the international perspective, usage data is attractive because it centralizes data pertinent to many different countries at a single entity, namely the AI provider. 
Notably, several frontier AI companies signed onto the New Delhi Frontier AI Commitments at the 2026 India AI Impact Summit to publish usage data reports.\footnote{The commitment to \textit{Advance Analysis on Real-World AI Usage} states that ``Participating organizations will work to enhance analysis regarding global
AI adoption for economic purposes \dots Publish --- by the next AI Summit --- statistical insights derived from anonymised,
aggregated and taxonomized usage data, either directly or (where relevant)
through contributions to international efforts.''}
For similar reasons, other private datasets such as payroll or payments information may help characterize frontier AI adoption while being naturally consolidated \citep[\eg][]{brynjolfsson2025canaries}.
On the other hand, many forms of traditional labor economics data hinge on government data collection infrastructure. 
Therefore, data quality and quantity may vary greatly across the world, generally leading to greater availability for advanced economies.
As our results show, frontier AI's impacts on advanced economies are unlikely to resemble impacts on developing nations.
Therefore, making the most of other data sources and sensibly imputing data in cases of missingness will be essential for characterizing the impacts of frontier AI on developing economies.

\paragraph{Limitations.}
The primary limitations of our work relate to limits of the underlying data we rely upon for occupational exposure estimates and national employment statistics.
First, we treat an occupation as identically exposed for every country.
While most work neglects within-country variation in an occupation (\eg software engineers in San Francisco may differ from those in New York City), we more strongly neglect cross-country intra-occupation variation (\eg software engineers in San Francisco may differ from those in New Delhi).
Second, we operate at the ISCO 2-digit level which collapses many distinct occupations into a single category.
For example, group 26 combines lawyers (2611) with musicians, singers, and composers (2652), even though their task bundles are qualitatively different. Third, the underlying occupational classification system (ISCO-08) was developed in 2008 and may not capture occupational emergence in the past two decades.
Looking forward, this extends to occupations that emerge due to frontier AI (\eg prompt engineers).
Fourth, we do not have complete coverage of the world's nations and we inherit the ways in which the ILO and World Bank handle national data for contentious regions of the world.
Perhaps most critically, we do not have national AI exposure estimates for China, Canada, and Saudi Arabia as a result.\footnote{We encourage future work to explore data imputation for these countries from domestic data surveys if data via the ILO is unavailable, though we refrained from doing this for analytic simplicity.}
\section*{Acknowledgements}
We thank Alex Spangher, Alexander Wan, Bharat Chandar, Diyi Yang, Dylan Clement, Erik Brynjolfsson, Sam Manning, Sarah Bana, Satya Krishna, Stephane Hatgis-Kessell, and Tom Cunningham for helpful discussion.

\bibliographystyle{unsrtnat}
\bibliography{main}

@article{eloundou2024gpts,
author = {Tyna Eloundou  and Sam Manning  and Pamela Mishkin  and Daniel Rock },
title = {GPTs are GPTs: Labor market impact potential of LLMs},
journal = {Science},
volume = {384},
number = {6702},
pages = {1306-1308},
year = {2024},
doi = {10.1126/science.adj0998},
URL = {https://www.science.org/doi/abs/10.1126/science.adj0998},
eprint = {https://www.science.org/doi/pdf/10.1126/science.adj0998}}

@techreport{narayanan2025ainormaltechnology,
  title       = {AI as Normal Technology: An alternative to the vision of AI as a potential superintelligence},
  author      = {Arvind Narayanan and Sayash Kapoor},
  institution = {Knight First Amendment Institute},
  year        = {2025},
  month       = apr,
  day         = {15},
  url         = {https://knightcolumbia.org/content/ai-as-normal-technology}
}

@techreport{acemoglu2024simple,
  author      = {Daron Acemoglu},
  title       = {The Simple Macroeconomics of AI},
  institution = {National Bureau of Economic Research},
  type        = {Working Paper},
  number      = {32487},
  year        = {2024},
  month       = may,
  doi         = {10.3386/w32487},
  url         = {https://www.nber.org/papers/w32487}
}

@misc{chandar2025ailabormarkets,
  author       = {Bharat Chandar},
  title        = {AI and Labor Markets: What We Know and Don't Know},
  year         = {2025},
  month        = oct,
  day          = {10},
  publisher    = {Stanford Digital Economy Lab},
  url          = {https://digitaleconomy.stanford.edu/news/ai-and-labor-markets-what-we-know-and-dont-know/},
  note         = {Accessed: 2026-04-09}
}

@misc{ipsos2025predictions,
  author       = {{Ipsos}},
  title        = {Ipsos Predictions Survey 2025: Positivity about how this year has gone highest since before the pandemic},
  year         = {2024},
  month        = dec,
  day          = {10},
  url          = {https://www.ipsos.com/en-us/ipsos-predictions-2025},
  note         = {Reports that globally 65\% expect AI-driven job losses and 43\% expect AI-driven job creation. Accessed: 2026-04-09}
}

@article{autor2003skill,
  author = {Autor, David H. and Levy, Frank and Murnane, Richard J.},
  title = {The Skill Content of Recent Technological Change: An Empirical Exploration},
  journal = {The Quarterly Journal of Economics},
  volume = {118},
  number = {4},
  pages = {1279--1333},
  year = {2003}
}

@incollection{acemoglu2011skills,
  author = {Acemoglu, Daron and Autor, David},
  title = {Skills, Tasks and Technologies: Implications for Employment and Earnings},
  booktitle = {Handbook of Labor Economics},
  editor = {Card, David and Ashenfelter, Orley},
  volume = {4},
  pages = {1043--1171},
  publisher = {Elsevier},
  year = {2011},
  doi = {10.1016/S0169-7218(11)02410-5}
}

@article{acemoglu2019automation,
  author = {Acemoglu, Daron and Restrepo, Pascual},
  title = {Automation and New Tasks: How Technology Displaces and Reinstates Labor},
  journal = {Journal of Economic Perspectives},
  volume = {33},
  number = {2},
  pages = {3--30},
  year = {2019},
  doi = {10.1257/jep.33.2.3}
}

@article{felten2018method,
  author = {Felten, Edward W. and Raj, Manav and Seamans, Robert},
  title = {A Method to Link Advances in Artificial Intelligence to Occupational Abilities},
  journal = {AEA Papers and Proceedings},
  volume = {108},
  pages = {54--57},
  year = {2018},
  doi = {10.1257/pandp.20181021}
}

@article{felten2021occupational,
  author = {Felten, Edward W. and Raj, Manav and Seamans, Robert},
  title = {Occupational, Industry, and Geographic Exposure to Artificial Intelligence: A Novel Dataset and Its Potential Uses},
  journal = {Strategic Management Journal},
  volume = {42},
  number = {12},
  pages = {2195--2217},
  year = {2021},
  doi = {10.1002/smj.3286}
}

@unpublished{webb2020impact,
  author = {Webb, Michael},
  title = {The Impact of Artificial Intelligence on the Labor Market},
  note = {Working paper},
  year = {2020},
  doi = {10.2139/ssrn.3482150}
}

@article{tolan2021measuring,
  author = {Tolan, Song{\"u}l and Pesole, Annarosa and Mart{\'i}nez-Plumed, Fernando and Fern{\'a}ndez-Mac{\'i}as, Enrique and Hern{\'a}ndez-Orallo, Jos{\'e} and G{\'o}mez, Emilia},
  title = {Measuring the Occupational Impact of AI: Tasks, Cognitive Abilities and AI Benchmarks},
  journal = {Journal of Artificial Intelligence Research},
  volume = {71},
  pages = {191--236},
  year = {2021},
  doi = {10.1613/jair.1.12647}
}

@article{felten2023occupational,
  author = {Felten, Edward W. and Raj, Manav and Seamans, Robert},
  title = {Occupational Heterogeneity in Exposure to Generative AI},
  journal = {SSRN Electronic Journal},
  year = {2023},
  doi = {10.2139/ssrn.4414065}
}

@techreport{pizzinelli2023labor,
  author = {Pizzinelli, Carlo and Panton, Augustus and Tavares, Marina M. and Cazzaniga, Mauro and Li, Longji},
  title = {Labor Market Exposure to AI: Cross-country Differences and Distributional Implications},
  institution = {International Monetary Fund},
  series = {IMF Working Paper},
  number = {WP/23/216},
  year = {2023}
}

@techreport{gmyrek2025refined,
  author = {Gmyrek, Pawel and Berg, Janine and Kami{\'n}ski, Karol and Konopczy{\'n}ski, Filip and {\L}adna, Agnieszka and Nafradi, Balint and Ros{\l}aniec, Konrad and Troszy{\'n}ski, Marek},
  title = {Generative AI and Jobs: A Refined Global Index of Occupational Exposure},
  institution = {International Labour Organization},
  series = {ILO Working Paper},
  number = {140},
  year = {2025},
  doi = {10.54394/HETP0387}
}

@online{massenkoffmccrory2026labor,
  author = {Massenkoff, Maxim and McCrory, Peter},
  title = {Labor Market Impacts of AI: A New Measure and Early Evidence},
  date = {2026-03-05},
  year = {2026},
  url = {https://www.anthropic.com/research/labor-market-impacts}
}

@techreport{demirer2026chaining,
  author = {Demirer, Mert and Horton, John J. and Immorlica, Nicole and Lucier, Brendan and Shahidi, Peyman},
  title = {Chaining Tasks, Redefining Work: A Theory of AI Automation},
  institution = {National Bureau of Economic Research},
  series = {NBER Working Paper},
  number = {34859},
  year = {2026},
  doi = {10.3386/w34859}
}

@techreport{gans2026oring,
  author = {Gans, Joshua S. and Goldfarb, Avi},
  title = {O-Ring Automation},
  institution = {National Bureau of Economic Research},
  series = {NBER Working Paper},
  number = {34639},
  year = {2026},
  doi = {10.3386/w34639}
}

@techreport{autor2025expertise,
 title = "Expertise",
 author = "Autor, David and Thompson, Neil",
 institution = "National Bureau of Economic Research",
 type = "Working Paper",
 series = "Working Paper Series",
 number = "33941",
 year = "2025",
 month = "June",
 doi = {10.3386/w33941},
 URL = "http://www.nber.org/papers/w33941",
}

@report{hai2025aiindex,
  author       = {{Stanford HAI}},
  title        = {Artificial Intelligence Index Report 2025},
  institution  = {Stanford University},
  year         = {2025},
  url          = {https://hai.stanford.edu/assets/files/hai_ai_index_report_2025.pdf},
  note         = {Accessed 2026-04-14}
}

@report{ipsos2025aimonitor,
  author       = {{Ipsos}},
  title        = {Ipsos AI Monitor 2025},
  year         = {2025},
  url          = {https://www.ipsos.com/sites/default/files/ct/publication/documents/2025-06/Ipsos-AI-Monitor-2025.pdf},
  note         = {Accessed 2026-04-14}
}

@misc{uk2025aiopportunities,
  author       = {{UK DSIT}},
  title        = {AI Opportunities Action Plan},
  howpublished = {Government of the United Kingdom},
  year         = {2025},
  month        = jan,
  url          = {https://www.gov.uk/government/publications/ai-opportunities-action-plan/ai-opportunities-action-plan},
  note         = {Accessed 2026-04-14}
}

@misc{canada2025sovereigncompute,
  author       = {{Canada ISED}},
  title        = {Canadian Sovereign AI Compute Strategy},
  year         = {2025},
  month        = oct,
  url          = {https://ised-isde.canada.ca/site/ised/en/canadian-sovereign-ai-compute-strategy},
  note         = {Accessed 2026-04-14}
}

@misc{ec2025aicontinent,
  author       = {{European Commission}},
  title        = {AI Continent Action Plan},
  year         = {2025},
  month        = apr,
  url          = {https://digital-strategy.ec.europa.eu/en/factpages/ai-continent-action-plan},
  note         = {Accessed 2026-04-14}
}

@misc{openai2025countries,
  author       = {{OpenAI}},
  title        = {Introducing OpenAI for Countries},
  year         = {2025},
  month        = may,
  url          = {https://openai.com/global-affairs/openai-for-countries/},
  note         = {Accessed 2026-04-14}
}

@report{unctad2025tir,
  author       = {{UN}},
  title        = {Technology and Innovation Report 2025: Inclusive Artificial Intelligence for Development},
  institution  = {United Nations},
  year         = {2025},
  url          = {https://unctad.org/system/files/official-document/tir2025_en.pdf},
  note         = {Accessed 2026-04-14}
}

@techreport{gmyrek2025exposure,
  author       = {Gmyrek, Pawe{\l} and Berg, Janine and Kami{\'n}ski, Karol and Konopczy{\'n}ski, Filip and {\L}adna, Agnieszka and Nafradi, Balint and Ros{\l}aniec, Konrad and Troszy{\'n}ski, Marek},
  title        = {Generative AI and Jobs: A Refined Global Index of Occupational Exposure},
  institution  = {International Labour Office},
  series       = {ILO Working Paper},
  number       = {140},
  address      = {Geneva},
  year         = {2025},
  doi          = {10.54394/HETP0387},
  url          = {https://doi.org/10.54394/HETP0387}
}

@techreport{cazzaniga2024genai,
  author      = {Cazzaniga, Mauro and Jaumotte, Florence and Li, Longji and Melina, Giovanni and Panton, Augustus J. and Pizzinelli, Carlo and Rockall, Emma J. and Tavares, Marina Mendes},
  title       = {{Gen-AI}: Artificial Intelligence and the Future of Work},
  institution = {International Monetary Fund},
  type        = {IMF Staff Discussion Note},
  number      = {SDN/2024/001},
  address     = {Washington, DC},
  year        = {2024},
  month       = jan,
  doi         = {10.5089/9798400262548.006},
  url         = {https://www.imf.org/en/publications/staff-discussion-notes/issues/2024/01/14/gen-ai-artificial-intelligence-and-the-future-of-work-542379}
}

@techreport{lewandowski2025workers,
  author      = {Lewandowski, Piotr and Mado{\'n}, Karol and Park, Albert},
  title       = {Workers' Exposure to {AI} Across Development},
  institution = {Instytut Badan Strukturalnych},
  type        = {IBS Working Paper},
  number      = {02/2025},
  year        = {2025},
  url         = {https://ibs.org.pl/wp-content/uploads/2025/03/Workers_AI_exposure_across_development_IBS_WP_202502.pdf}
}

@techreport{gmyrek2026disruption,
  author      = {Gmyrek, Pawe{\l} and Viollaz, Mariana and Winkler, Hernan},
  title       = {Disruption without Dividend?: How the Digital Divide and Task Differences Split {GenAI}'s Global Impact},
  institution = {International Labour Organization},
  type        = {ILO Working Paper},
  number      = {166},
  address     = {Geneva},
  year        = {2026},
  doi         = {10.54394/00033147},
  url         = {https://www.ilo.org/sites/default/files/2026-03/wp166_web.pdf}
}

@article{brynjolfsson2025generative,
  author  = {Brynjolfsson, Erik and Li, Danielle and Raymond, Lindsey R.},
  title   = {Generative AI at Work},
  journal = {The Quarterly Journal of Economics},
  volume  = {140},
  number  = {2},
  pages   = {889--942},
  year    = {2025},
  doi     = {10.1093/qje/qjae044},
  url     = {https://academic.oup.com/qje/article/140/2/889/7990658}
}

@article{dellacqua2026jagged,
  author  = {Dell'Acqua, Fabrizio and McFowland III, Edward and Mollick, Ethan and Lifshitz, Hila and Kellogg, Katherine C. and Rajendran, Saran and Krayer, Lisa and Candelon, Fran{\c{c}}ois and Lakhani, Karim R.},
  title   = {Navigating the Jagged Technological Frontier: Field Experimental Evidence of the Effects of Artificial Intelligence on Knowledge Worker Productivity and Quality},
  journal = {Organization Science},
  volume  = {37},
  number  = {2},
  pages   = {403--423},
  year    = {2026},
  doi     = {10.1287/orsc.2025.21838},
  url     = {https://pubsonline.informs.org/doi/10.1287/orsc.2025.21838}
}

@misc{ilostat2026,
  author       = {{International Labour Organization}},
  title        = {ILOSTAT Database},
  year         = {2026},
  url          = {https://ilostat.ilo.org/data/},
  note         = {Accessed 2026-04-14}
}

@techreport{ilo2024careparticipation,
  author      = {{International Labour Organization}},
  title       = {The Impact of Care Responsibilities on Women's Labour Force Participation},
  institution = {International Labour Organization},
  type        = {Statistical Brief},
  year        = {2024},
  month       = oct,
  doi         = {10.54394/LPTT5569},
  url         = {https://www.ilo.org/sites/default/files/2024-10/GEDI-STAT%20brief_formatted_28.10.24_final.pdf}
}

@article{blau2017genderwagegap,
  author  = {Blau, Francine D. and Kahn, Lawrence M.},
  title   = {The Gender Wage Gap: Extent, Trends, and Explanations},
  journal = {Journal of Economic Literature},
  volume  = {55},
  number  = {3},
  pages   = {789--865},
  year    = {2017},
  doi     = {10.1257/jel.20160995},
  url     = {https://www.aeaweb.org/articles?id=10.1257/jel.20160995}
}

@article{humlum2025unequal,
  author  = {Humlum, Anders and Vestergaard, Emilie},
  title   = {The Unequal Adoption of ChatGPT Exacerbates Existing Inequalities Among Workers},
  journal = {Proceedings of the National Academy of Sciences of the United States of America},
  volume  = {122},
  number  = {1},
  pages   = {e2414972121},
  year    = {2025},
  doi     = {10.1073/pnas.2414972121},
  url     = {https://doi.org/10.1073/pnas.2414972121}
}

@techreport{otis2024globalgendergapsgenai,
  author      = {Otis, Nicholas G. and Delecourt, Sol{\`e}ne and Cranney, Katelyn and Koning, Rembrand},
  title       = {Global Evidence on Gender Gaps and Generative AI},
  institution = {Harvard Business School},
  type        = {Working Paper},
  number      = {25-023},
  year        = {2024},
  url         = {https://www.hbs.edu/ris/download.aspx?name=25023.pdf}
}

@techreport{mcclain2025publicexpertsai,
  author      = {McClain, Colleen and Kennedy, Brian and Gottfried, Jeffrey and Anderson, Monica and Pasquini, Giancarlo},
  title       = {How the U.S. Public and AI Experts View Artificial Intelligence},
  institution = {Pew Research Center},
  year        = {2025},
  month       = apr,
  day         = {3},
  url         = {https://www.pewresearch.org/internet/2025/04/03/how-the-us-public-and-ai-experts-view-artificial-intelligence/}
}

@techreport{pew2025globalai,
  author      = {{Pew Research Center}},
  title       = {How People Around the World View AI},
  institution = {Pew Research Center},
  year        = {2025},
  month       = oct,
  day         = {15},
  url         = {https://www.pewresearch.org/global/2025/10/15/how-people-around-the-world-view-ai/}
}

@techreport{kpmg2025aiglobal,
  author      = {Gillespie, Nicole and Lockey, Steve and Ward, T. and Macdade, A. and Hassed, G.},
  title       = {Trust, attitudes and use of artificial intelligence: A global study 2025},
  institution = {The University of Melbourne and KPMG},
  year        = {2025},
  doi         = {10.26188/28822919},
  url         = {https://figshare.unimelb.edu.au/articles/report/Trust_attitudes_and_use_of_artificial_intelligence_A_global_study_2025/28822919},
  note        = {Accessed 2026-04-17}
}

@techreport{manning2026adaptable,
 title = "How Adaptable Are American Workers to AI-Induced Job Displacement?",
 author = "Manning, Sam J and Aguirre, Tomás",
 institution = "National Bureau of Economic Research",
 type = "Working Paper",
 series = "Working Paper Series",
 number = "34705",
 year = "2026",
 month = "January",
 doi = {10.3386/w34705},
 URL = "http://www.nber.org/papers/w34705",
}

@article{tomlinson2025workingwithai,
  title   = {Working with AI: Measuring the Applicability of Generative AI to Occupations},
  author  = {Tomlinson, Kiran and Jaffe, Sonia and Wang, Will and Counts, Scott and Suri, Siddharth},
  journal = {arXiv preprint arXiv:2507.07935},
  year    = {2025},
  doi     = {10.48550/arXiv.2507.07935},
  url     = {https://arxiv.org/abs/2507.07935}
}

@techreport{chatterji2025howpeopleusechatgpt,
  title       = {How People Use ChatGPT},
  author      = {Chatterji, Aaron and Cunningham, Thomas and Deming, David J. and Hitzig, Zoe and Ong, Christopher and Shan, Carl Yan and Wadman, Kevin},
  institution = {National Bureau of Economic Research},
  type        = {Working Paper},
  number      = {34255},
  year        = {2025},
  month       = sep,
  doi         = {10.3386/w34255},
  url         = {https://www.nber.org/papers/w34255}
}

@article{handa2025economictasks,
  title   = {Which Economic Tasks Are Performed with AI? Evidence from Millions of Claude Conversations},
  author  = {Handa, Kunal and Tamkin, Alex and McCain, Miles and Huang, Saffron and Durmus, Esin and Heck, Sarah and Mueller, Jared and Hong, Jerry and Ritchie, Stuart and Belonax, Tim and Troy, Kevin K. and Amodei, Dario and Kaplan, Jared and Clark, Jack and Ganguli, Deep},
  journal = {arXiv preprint arXiv:2503.04761},
  year    = {2025},
  doi     = {10.48550/arXiv.2503.04761},
  url     = {https://arxiv.org/abs/2503.04761}
}

@techreport{brynjolfsson2025canaries,
  title       = {Canaries in the Coal Mine? Six Facts about the Recent Employment Effects of Artificial Intelligence},
  author      = {Brynjolfsson, Erik and Chandar, Bharat and Chen, Ruyu},
  institution = {Stanford Digital Economy Lab},
  year        = {2025},
  month       = nov,
  url         = {https://digitaleconomy.stanford.edu/app/uploads/2025/11/CanariesintheCoalMine_Nov25.pdf}
}

@online{appelmccrorytamkin2025geoapi,
  author = {Appel, Ruth and McCrory, Peter and Tamkin, Alex and Stern, Michael and McCain, Miles and Neylon, Tyler},
  title  = {Anthropic Economic Index report: Uneven geographic and enterprise AI adoption},
  year   = {2025},
  month  = sep,
  day    = {15},
  url    = {https://www.anthropic.com/research/anthropic-economic-index-september-2025-report}
}

@techreport{openai2026signalsglobal,
  title       = {OpenAI Signals: Measuring AI Adoption, Protecting Privacy, and Empowering Decisions},
  author = {OpenAI},
  year        = {2026},
  month       = feb,
  note        = {Global affairs report},
  url         = {https://cdn.openai.com/signals/openai-signals-global-report.pdf}
}

@techreport{tucker2026qwi,
  author = {Tucker, Lee C.},
  title = {You're (not) hired: Artificial intelligence and early career hiring in the Quarterly Workforce Indicators},
  year = {2026},
  month = apr,
  day = {17},
  url = {http://leetucker.net/docs/Youre_not_hired_Tucker_20260417.pdf}
}

@online{imas2026howwillai,
  author       = {Imas, Alex and Shukla, Soumitra},
  title        = {How Will AI-driven Automation Actually Affect Jobs?},
  subtitle     = {The economics of AI exposure and job displacement},
  year         = {2026},
  month        = mar,
  day          = {23},
  organization = {Ghosts of Electricity},
  url          = {https://aleximas.substack.com/p/how-will-ai-driven-automation-actually}
}

@misc{worldbank_knomad_bre,
  author       = {{World Bank}},
  title        = {Bilateral Remittance Estimates using Migrant Stocks, Host Country Incomes, and Origin Country Incomes ({US\$} million)},
  year         = {2026},
  howpublished = {\url{https://data360.worldbank.org/en/indicator/WB_KNOMAD_BRE}},
  note         = {KNOMAD/World Bank Data 360 indicator, accessed April 19, 2026}
}

@misc{worldbank_knomad_mri,
  author       = {{World Bank}},
  title        = {Remittance inflows ({US\$} million)},
  year         = {2026},
  howpublished = {\url{https://data360.worldbank.org/en/indicator/WB_KNOMAD_MRI}},
  note         = {KNOMAD/World Bank Data 360 indicator, accessed April 19, 2026}
}

@misc{ratha2022bilateralmatrix,
  author       = {Ratha, Dilip and Plaza, Sonia and Kim, Eung Ju},
  title        = {Bilateral Remittance Matrix (new)},
  year         = {2022},
  month        = dec,
  howpublished = {\url{https://blogs.worldbank.org/en/peoplemove/bilateral-remittance-matrix-new}},
  note         = {World Bank People Move blog, December 19, 2022}
}

@misc{eurostat2024financeinsurance,
  author       = {{Eurostat}},
  title        = {4.9 million employed in finance and insurance sector},
  year         = {2024},
  month        = jun,
  day          = {10},
  howpublished = {\url{https://ec.europa.eu/eurostat/web/products-eurostat-news/w/ddn-20240610-2}},
  note         = {Accessed 2026-04-20}
}

@techreport{mom2026labourforcesingapore,
  author       = {{Ministry of Manpower, Singapore}},
  title        = {Labour Force in Singapore 2025},
  institution  = {{Ministry of Manpower, Singapore}},
  year         = {2026},
  month        = feb,
  howpublished = {\url{https://stats.mom.gov.sg/iMAS_PdfLibrary/mrsd_2025Labourforce.pdf}},
  note         = {Accessed 2026-04-20}
}

@techreport{worldbank2024burundiyouth,
  author       = {{World Bank}},
  title        = {Burundi: Supporting youth-at-risk through the creative sector (P505887) Project Information Document (PID)},
  institution  = {{World Bank}},
  year         = {2024},
  month        = nov,
  day          = {15},
  number       = {PIDDC00690},
  howpublished = {\url{https://documents1.worldbank.org/curated/en/099121124070521165/pdf/P505887187295c01f1b4321f9f82816289a.pdf}},
  note         = {Accessed 2026-04-20}
}

@techreport{psa2024womenmenwork,
  author       = {{Philippine Statistics Authority}},
  title        = {2024 Fact Sheet on Women and Men in the Philippines: Work and Economic Participation},
  institution  = {{Philippine Statistics Authority}},
  year         = {2024},
  howpublished = {\url{https://psa.gov.ph/sites/default/files/phdsd/Work%20and%20Economic%20Participation%202024%20Fact%20Sheet%20on%20Women%20and%20Men_0.pdf}},
  note         = {Accessed 2026-04-20}
}

@techreport{statin2020jamaicanlabourmarket,
  author       = {{Statistical Institute of Jamaica}},
  title        = {Jamaican Labour Market: Impact of COVID-19},
  institution  = {{Statistical Institute of Jamaica}},
  year         = {2020},
  month        = jul,
  day          = {19},
  howpublished = {\url{https://statinja.gov.jm/covidPDF/Jamaican%20Labour%20Market%20Impact%20of%20COVID-19.pdf}},
  note         = {Accessed 2026-04-20}
}

@techreport{pbs2026lfs202425,
  author       = {{Pakistan Bureau of Statistics}},
  title        = {Labour Force Survey 2024--25},
  institution  = {{Pakistan Bureau of Statistics}},
  year         = {2026},
  howpublished = {\url{https://www.pbs.gov.pk/wp-content/uploads/2020/07/LFS-2024-25-Annual-Report.pdf}},
  note         = {Accessed 2026-04-20}
}

@techreport{mospi2023womenmenindia,
  author       = {{Ministry of Statistics and Programme Implementation}},
  title        = {Women and Men in India 2023: Participation in Economy},
  institution  = {{Ministry of Statistics and Programme Implementation, Government of India}},
  year         = {2023},
  howpublished = {\url{https://www.mospi.gov.in/sites/default/files/reports_and_publication/statistical_publication/Women_Men/mw23/Participation_in_economy.pdf}},
  note         = {Accessed 2026-04-20}
}

@misc{onet2026datadictionary302,
  author       = {{O*NET Resource Center}},
  title        = {O*NET 30.2 Data Dictionary},
  year         = {2026},
  month        = apr,
  day          = {14},
  howpublished = {\url{https://www.onetcenter.org/dictionary/30.2/text/}},
  note         = {Site updated 2026-04-14; accessed 2026-04-20}
}

@techreport{microsoft2026aidiffusion,
  title       = {Global AI Adoption in 2025: A Widening Digital Divide},
  author      = {{Microsoft AI Economy Institute}},
  institution = {Microsoft},
  year        = {2026},
  url         = {https://www.microsoft.com/en-us/corporate-responsibility/topics/AI-Economy-Institute/reports/Global-AI-Adoption-2025/}
}

@online{massenkoffhuang2026economics,
  author = {Maxim Massenkoff and Saffron Huang},
  title = {What 81,000 People Told Us About the Economics of AI},
  year = {2026},
  date = {2026-04-22},
  url = {https://www.anthropic.com/research/81k-economics}
}

@online{manning2025DOL,
  author = {Sam Manning},
  title = {Understanding AI's Labor Market Impacts},
  year = {2025},
  url = {https://cdn.sanity.io/files/d8lrla4f/staging/6c5cbe4cbe50f97034173d7f4b55ae1f1377b655.pdf}
}

@techreport{hosseinimaasoum2026generative,
  author      = {Hosseini Maasoum, Seyed Mahdi and Lichtinger, Guy},
  title       = {Generative AI, Expertise, and Effective Labor Supply},
  institution = {SSRN},
  year        = {2026},
  month       = jan,
  url         = {https://ssrn.com/abstract=6059674},
  doi         = {10.2139/ssrn.6059674}
}

@online{anthropic2026aeiv5,
        author = {Maxim Massenkoff and Eva Lyubich and Peter McCrory and Ruth Appel and Ryan Heller},
        title = {Anthropic Economic Index report: Learning curves},
        date = {2026-03-24},
        year = {2026},
        url = {https://www.anthropic.com/research/economic-index-march-2026-report},
}

@misc{microsoft2026aidiffusiondata,
  title        = {AI Diffusion Report Data},
  author       = {{Microsoft AI Economy Institute}},
  year         = {2026},
  howpublished = {\url{https://github.com/microsoft/ai-diffusion-report}},
  note         = {Q1 2026 update; MIT licensed data repository; accessed May 19, 2026}
}

\newpage
\appendix
\section{Data}
\label{app:data}
\subsection{ILO employment}
\label{app:data_ilo}

\begin{table}[ht]
\centering
\caption{Country coverage in the ILOSTAT employment pipeline}
\label{tab:sample_coverage}
\begin{tabular}{lr}
\toprule
Status & Countries \\
\midrule
Included in baseline sample (medium/high reliability) & 141 \\
Observed but excluded for low reliability & 6 \\
Excluded: TOTAL/NEC only & 2 \\
Excluded: ISCO-88 only & 9 \\
Excluded: major-group only & 9 \\
\bottomrule
\end{tabular}
\vspace{0.4em}\par\footnotesize{Baseline sample counts reflect the current merged ILO refresh and include countries with medium or high reliability only.}
\end{table}

We use employment data published by the International Labour Organization (ILO), taking the most recent annual country observation available in the local ILO mirror accessed in April 2026.
The total-employment country-years in the underlying panel range from 2009 to 2025, with most observations from 2024 or 2025.
The employment data used in the main analysis are reported at the 2-digit (sub-major group) level of ISCO-08, yielding up to 40 occupation groups per country once armed-forces, aggregate, and residual categories are excluded.
The three excluded armed-forces sub-major groups are ISCO-08 codes \texttt{01}, \texttt{02}, and \texttt{03}; we also drop aggregate \texttt{TOTAL} rows and residual \texttt{X} rows rather than redistributing them.
In the ILO mirror \texttt{EMP\_TEMP\_SEX\_OC2\_NB\_A}, 113 countries report nonzero employment in the armed-forces groups.
Among those countries, the excluded armed-forces share is usually small (median 0.42\%) but reaches 5.11\% in Iraq, 4.61\% in Palestine, 4.46\% in Lebanon, and 2.95\% in Sudan.
While the data nominally covers 167 nations, we analyze the 141 where the ILO ascribes greater confidence to the underlying data quality and data collecting procedures. Gender-disaggregated exposure estimates use the same ILO occupational-employment pipeline restricted to sex-disaggregated country-occupation cells.
For the main white-collar measure, we classify ISCO-08 major groups \texttt{1}--\texttt{4} (managers, professionals, technicians and associate professionals, and clerical support workers) as white-collar work.
Sales occupations are treated separately in the mechanism analysis because ISCO-08 code \texttt{52} is a boundary case: including sales with white-collar work substantially increases the explanatory power of the simple labor-composition split.

\subsection{ISCO exposure estimates}
\label{app:data_gmyrek}

\begin{table}[H]
\centering
\caption{Hierarchy of occupational exposure estimates built from Gmyrek et al. (2025)}
\label{tab:gmyrek_hierarchy_summary}
\renewcommand{\arraystretch}{1.08}

\footnotesize
\setlength{\tabcolsep}{4pt}
\renewcommand{\arraystretch}{0.95}

\begin{tabular}{>{\raggedright\arraybackslash}p{3.5cm}
                >{\centering\arraybackslash}p{0.9cm}
                >{\raggedright\arraybackslash}p{4.2cm}
                >{\raggedright\arraybackslash}p{3.9cm}}
\toprule
Level & Count & Aggregation rule & Use in this paper \\
\midrule
Task-by-occupation rows in the Gmyrek workbook
& 3,265
& Exposure score assigned to each task row within a 4-digit ISCO-08 occupation
& Raw exposure source layer \\

4-digit occupations (ISCO-08)
& 427
& Mean task-level exposure score within each 4-digit occupation
& Occupation-level exposure estimates \\

2-digit occupations
& 40
& Mean 4-digit occupation score within each 2-digit ISCO-08 group
& Matched to ILO employment shares in the main cross-country analysis \\

1-digit occupation groups
& 9
& Aggregation of 2-digit occupations to ISCO major groups
& Descriptive summaries and wage-weighted extensions \\
\bottomrule
\end{tabular}

\vspace{0.35em}
\begin{minipage}{0.98\linewidth}
\footnotesize\textit{Note:} Counts are based on the Gmyrek et al. (2025) exposure workbook used in our pipeline. The workbook contains 3,219 predicted rows and 46 reconciled rows.
\end{minipage}
\end{table}

We use the occupational exposure estimates for ISCO-08 \citep{gmyrek2025refined}.
We describe these estimates in terms of their relationship to the ISCO-08 multi-level hierarchy in \autoref{tab:gmyrek_hierarchy_summary}.
Since we only have employment data at the 2-digit level of ISCO-08, we average the occupational exposure estimates for all occupations within the same 2-digit category.
Of the 436 ISCO-08 4-digit occupations, the Gmyrek et al. data score 427.
The nine unscored 4-digit codes are six ``not elsewhere classified'' residual codes (\texttt{1439}, \texttt{3139}, \texttt{3435}, \texttt{5249}, \texttt{7319}, \texttt{8189}) and three armed-forces codes (\texttt{0110}, \texttt{0210}, \texttt{0310}).
We do not impute exposure scores for these unscored residual or armed-forces occupations.

\subsection{World Bank national statistics}
\label{app:data_worldbank}
We merge the occupational exposure panel with World Bank national statistics used in the descriptive and regression analyses, including GNI per capita in PPP terms, internet penetration, population, and remittance indicators.
GNI and internet penetration enter the appendix predictor regressions as country-level development and infrastructure controls.

\subsection{World Bank KNOMAD remittance}
\label{app:data_knomad}
We use World Bank WDI data on personal remittances received to measure remittance inflows in current U.S. dollars and as a share of GDP, using reported WDI values directly where available.
We use the KNOMAD bilateral remittance matrix (\texttt{WB.KNOMAD.BRE}) to estimate source-country shares for remittance-receiving countries, again using the reported matrix values directly; the latest bilateral matrix available in our data is for 2021.
The remittance examples therefore combine the latest available WDI remittance totals or remittance-to-GDP shares, generally 2024, with 2021 KNOMAD bilateral source-country shares.
We do not impute missing bilateral corridors, extrapolate bilateral shares to later years, or assign synthetic exposure values.

\subsection{Usage data statistics}
\label{app:data_usage}
The observed adoption analysis uses three external usage statistics: Anthropic Economic Index country-level Claude usage, OpenAI Signals country rankings for population-normalized ChatGPT use, and Microsoft AI Diffusion country-level generative AI adoption rates \citep{appelmccrorytamkin2025geoapi,handa2025economictasks,openai2026signalsglobal,microsoft2026aidiffusion,microsoft2026aidiffusiondata}.
We match these external country statistics to the 141-country exposure panel using ISO-3 country codes where publishers provide them and otherwise using unambiguous country or economy names.
We do not adjudicate contested political status or reassign entities across source definitions: labels such as Taiwan, China, Korea, or other country/economy groupings are retained as reported by the source, and an observation is used only when it maps unambiguously to an entity in the exposure panel.
If a source reports an entity that does not cleanly match the exposure panel, we omit it rather than imputing or reallocating it.
Each adoption regression uses the complete-case country sample available for the corresponding outcome and covariates.
The resulting complete-case samples are 114 countries for Anthropic Claude usage, 88 for OpenAI Signals, and 106 for Microsoft AI Diffusion.

\subsection{Reproducibility and asset provenance}
\label{app:reproducibility}
The anonymous release for reviewers has two parts.
The dataset release contains the derived measured CSV tables used in the paper, Croissant metadata, a data dictionary, and a source-data manifest.
The reviewer code package contains validation scripts, figure and table builders, and two clean notebooks (\texttt{01\_dataset\_tour.ipynb} and \texttt{02\_reproduce\_main\_results.ipynb}) that reproduce paper-facing values, regression tables, and release-supported figures from the released derived tables where redistribution permits.
The analysis consists of deterministic data validation, aggregation, crosswalks, plotting, and ordinary least squares regressions run in Python on CPU; no model training or GPU computation is required.
The released derived tables freeze the values used in this submission.
Raw-source rebuilds are deterministic conditional on the source snapshots listed in \autoref{tab:source_data_provenance}, but some upstream ILO and World Bank mirrors may be revised over time; for this reason, the reviewer workflow does not require live API pulls and instead reproduces results from the released tables.
Where source data cannot be redistributed under the provider's terms, the release records the original source, access date or version, redistribution status, and release tables supported in the source-data manifest.
Raw third-party files and non-anonymized working notebooks are not redistributed; the Microsoft AI Diffusion country/economy adoption table is included from Microsoft's official MIT-licensed GitHub data release after ISO alpha-3 matching.

\begin{table}[ht]
\centering
\caption{Source data and provenance for the main analyses.}
\label{tab:source_data_provenance}
\footnotesize
\setlength{\tabcolsep}{3pt}
\begin{tabular}{p{0.22\linewidth}p{0.27\linewidth}p{0.18\linewidth}p{0.23\linewidth}}
\toprule
Asset & Use in paper & Version or date & Credit and license/terms source \\
\midrule
ILOSTAT employment and wage statistics & National occupational employment and wage composition & Most recent annual country statistics; accessed 2026-04-14 & \citet{ilostat2026}; \url{https://ilostat.ilo.org/data/} \\
Gmyrek et al. ISCO exposure estimates & Primary occupation-level exposure scores & 2025 ILO Working Paper 140 release & \citet{gmyrek2025refined} \\
World Bank WDI and KNOMAD & GNI, remittance totals, and bilateral remittance shares & WDI 2024 where available; KNOMAD bilateral matrix latest year 2021 & \citet{worldbank_knomad_bre,worldbank_knomad_mri}; \url{https://databank.worldbank.org/} \\
Anthropic Economic Index & Country-level Claude usage validation & Country usage release used in \autoref{fig:observed_outcomes_vs_exposure} & \citet{anthropic2026aeiv5} \\
OpenAI Signals & Population-normalized ChatGPT usage rank validation & Calendar-year 2025 country ranks & \citet{openai2026signalsglobal} \\
Microsoft AI Diffusion & Country-level generative AI adoption validation & Q1 2026 GitHub data update & \citet{microsoft2026aidiffusion,microsoft2026aidiffusiondata}; MIT license at \url{https://github.com/microsoft/ai-diffusion-report} \\
O*NET 30.2 and alternative exposure estimates & Occupational-property and robustness analyses & O*NET 30.2; source releases cited in appendix & \citet{onet2026datadictionary302,eloundou2024gpts,hosseinimaasoum2026generative} \\
\bottomrule
\end{tabular}
\end{table}
\section{Exposure Analysis}
\label{app:exposure}

\subsection{Alternative Occupational Exposure Estimates}
\label{app:alternatives}
Throughout our work, we use the occupational exposure estimates of \citet{gmyrek2025refined} because they are naturally designed for the ISCO-08 occupational categories.
However, other exposure estimates are more popular in the broader economics of AI literature, which often derive from the U.S. Department of Labor such as the O*NET work taxonomy and the SOC occupational list.
Via a crosswalk that aligns the DOL resources and the SOC list with the ILO resources and the ISCO list, we can compare the occupational exposure estimates we use to other popular choices.

We compare our primary Gmyrek et al. exposure scores to two prominent alternatives: the LLM-task exposure measure from \citet{eloundou2024gpts} and the occupational exposure measure from \citet{hosseinimaasoum2026generative}.
Because these alternatives are defined over O*NET/SOC occupations rather than ISCO-08 occupations, we use the SOC-to-ISCO crosswalk described above and aggregate scores to ISCO-08 2-digit occupations before recomputing national exposure scores.\footnote{The robustness pipeline maps O*NET-SOC 8-digit occupations to SOC-2018 6-digit occupations, then to SOC-2010 6-digit occupations using the BLS 2018 reclassification crosswalk, then to ISCO-08 4-digit occupations using the BLS 2012 SOC-ISCO crosswalk, and finally to ISCO-08 2-digit occupations. At each merge, we take unweighted means across crosswalk rows. The BLS SOC-ISCO crosswalk is many-to-many and includes partial-overlap flags, which we do not weight separately. As a sensitivity check, assigning each SOC-to-ISCO mapping a part-share weight of $1/k$ for a SOC code that maps to $k$ ISCO codes changes the Gmyrek-vs-Eloundou Pearson correlation from $0.954$ to $0.957$ and the Spearman correlation from $0.950$ to $0.955$, leaving the robustness conclusion unchanged.}

The estimates are highly correlated at the occupational and national levels (see \autoref{fig:alternative_exposure_estimates}).
At the occupation level, the resulting measures are highly correlated with the \citet{gmyrek2025exposure} scores: Pearson correlations are $r=0.954$ for \citet{eloundou2024gpts} and $r=0.948$ for \citet{hosseinimaasoum2026generative}.
At the national level, rankings are even more stable, with Spearman correlations of $\rho=0.993$ and $\rho=0.990$, respectively.
The main disagreements are concentrated in a small number of occupations, such as ICT professionals and assemblers, but these differences do not materially change the cross-country ranking of national exposure.

\begin{figure}[ht]
    \centering
    \includegraphics[width=\linewidth]{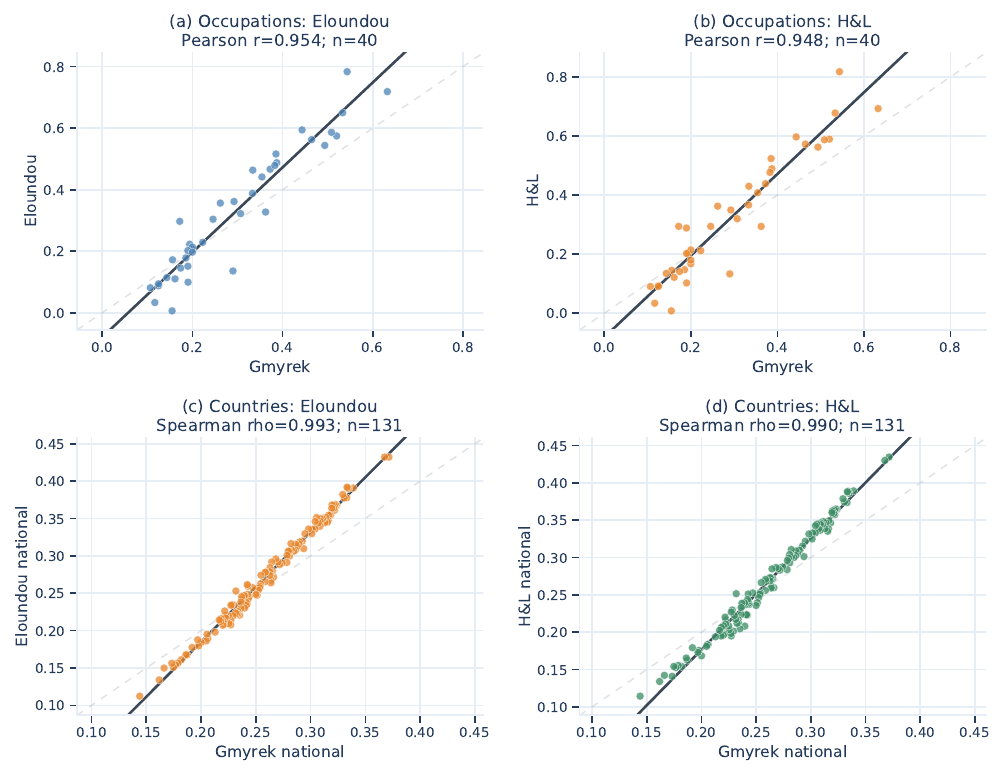}
    \caption{\textbf{National exposure estimates are robust to alternative occupational exposure measures.}
    The top row compares ISCO-08 2-digit occupational exposure scores. The bottom row compares national exposure scores after reweighting each occupational index by country-level occupational employment. The alternative indices differ in their exact task definitions and source occupational taxonomies, but they produce very similar occupational scores and highly stable national rankings.}
    \label{fig:alternative_exposure_estimates}
\end{figure}

\subsection{National Exposure Distribution and Country Scores}
\label{app:national_exposure_scores}

The main text visualizes national AI exposure geographically in \autoref{fig:national_exposure_distribution}.
For completeness, \autoref{fig:appendix_national_exposure_distribution} shows the corresponding distributional view, including the least- and most-exposed countries among countries with at least one million workers.
\autoref{tab:national_exposure_scores} reports the full 141-country exposure panel used in the map.

\begin{figure}[ht]
    \centering
    \includegraphics[width=\linewidth]{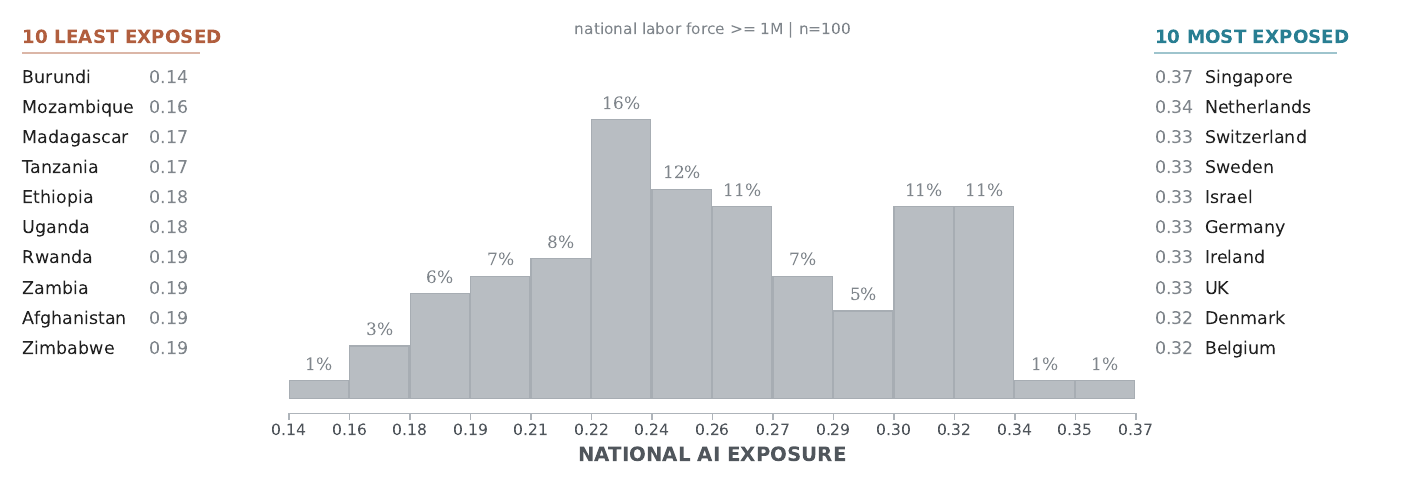}
    \caption{\textbf{Distribution of national AI exposure estimates.}
    The side panels list the ten least- and most-exposed countries subject to the labor-force threshold used for visual readability; the histogram summarizes the full measured sample.}
    \label{fig:appendix_national_exposure_distribution}
\end{figure}

\begin{table}[ht]
\centering
\caption{National AI exposure estimates for all measured countries.}
\label{tab:national_exposure_scores}
\scriptsize
\setlength{\tabcolsep}{4pt}
\resizebox{\linewidth}{!}{%
\begin{tabular}{lrlrlr}
\toprule
Country & ${n}$ & Country & ${n}$ & Country & ${n}$ \\
\midrule
Afghanistan & 0.189 & Germany & 0.333 & Norway & 0.320 \\
Albania & 0.230 & Ghana & 0.220 & Pakistan & 0.225 \\
Angola & 0.206 & Greece & 0.315 & Palau & 0.295 \\
Argentina & 0.294 & Guatemala & 0.234 & Palestine (State of) & 0.253 \\
Australia & 0.322 & Guinea & 0.227 & Panama & 0.232 \\
Austria & 0.321 & Guinea-Bissau & 0.179 & Papua New Guinea & 0.236 \\
Bahamas & 0.282 & Guyana & 0.247 & Peru & 0.243 \\
Bangladesh & 0.218 & Honduras & 0.231 & Philippines & 0.252 \\
Barbados & 0.286 & Hungary & 0.309 & Poland & 0.317 \\
Belarus & 0.282 & India & 0.217 & Portugal & 0.314 \\
Belgium & 0.323 & Indonesia & 0.242 & Romania & 0.272 \\
Belize & 0.242 & Iran (Islamic Republic of) & 0.254 & Russian Federation & 0.298 \\
Benin & 0.227 & Iraq & 0.243 & Rwanda & 0.186 \\
Bhutan & 0.242 & Ireland & 0.331 & Samoa & 0.239 \\
Bolivia (Plurinational State of) & 0.241 & Israel & 0.333 & Sao Tome and Principe & 0.230 \\
Bosnia and Herzegovina & 0.263 & Italy & 0.313 & Serbia & 0.285 \\
Botswana & 0.228 & Jamaica & 0.266 & Sierra Leone & 0.205 \\
Brazil & 0.278 & Japan & 0.316 & Singapore & 0.368 \\
Brunei Darussalam & 0.307 & Jordan & 0.255 & Slovenia & 0.304 \\
Bulgaria & 0.292 & Kenya & 0.199 & Solomon Islands & 0.199 \\
Burkina Faso & 0.198 & Kiribati & 0.227 & Somalia & 0.259 \\
Burundi & 0.144 & Kosovo & 0.288 & Spain & 0.301 \\
Cabo Verde & 0.220 & Kyrgyzstan & 0.240 & Sri Lanka & 0.237 \\
Cambodia & 0.221 & Lao People's Democratic Republic & 0.205 & Sudan & 0.213 \\
Chile & 0.279 & Latvia & 0.304 & Suriname & 0.270 \\
Colombia & 0.263 & Lebanon & 0.264 & Sweden & 0.334 \\
Congo, Democratic Republic of the & 0.197 & Lesotho & 0.175 & Switzerland & 0.335 \\
Cook Islands & 0.290 & Lithuania & 0.309 & Tajikistan & 0.222 \\
Costa Rica & 0.274 & Luxembourg & 0.371 & Tanzania, United Republic of & 0.173 \\
Croatia & 0.311 & Madagascar & 0.166 & Thailand & 0.250 \\
Cuba & 0.265 & Malawi & 0.223 & Timor-Leste & 0.208 \\
Cyprus & 0.322 & Maldives & 0.280 & Togo & 0.235 \\
Czechia & 0.302 & Mali & 0.200 & Tokelau & 0.273 \\
Côte d'Ivoire & 0.240 & Marshall Islands & 0.268 & Tonga & 0.259 \\
Denmark & 0.324 & Mauritius & 0.289 & Tunisia & 0.236 \\
Dominican Republic & 0.264 & Mexico & 0.264 & Tuvalu & 0.305 \\
Ecuador & 0.222 & Micronesia (Federated States of) & 0.219 & Türkiye & 0.265 \\
Egypt & 0.254 & Mongolia & 0.263 & Uganda & 0.182 \\
El Salvador & 0.251 & Montenegro & 0.318 & United Arab Emirates & 0.281 \\
Estonia & 0.311 & Montserrat & 0.290 & United Kingdom of Great Britain and Northern Ireland & 0.329 \\
Eswatini & 0.244 & Mozambique & 0.162 & United States of America & 0.323 \\
Ethiopia & 0.177 & Myanmar & 0.229 & Uruguay & 0.281 \\
Fiji & 0.259 & Nepal & 0.233 & Vanuatu & 0.218 \\
Finland & 0.309 & Netherlands & 0.339 & Viet Nam & 0.225 \\
France & 0.319 & Niger & 0.241 & Wallis and Futuna & 0.291 \\
Gambia & 0.234 & Niue & 0.314 & Zambia & 0.187 \\
Georgia & 0.237 & North Macedonia & 0.286 & Zimbabwe & 0.192 \\
\bottomrule
\end{tabular}
}
\end{table}

\clearpage

\subsection{Gender Differences in National AI Exposure}
\label{app:gender_exposure}

\autoref{fig:appendix_gender_exposure_gap} visualizes the country-level gender exposure gap discussed in the main text.
The x-axis reports the female-minus-male exposure gap as a percentage of total national exposure, so positive values indicate that female exposure exceeds male exposure.
The pattern is broad rather than driven by a handful of countries: women are more exposed than men in most measured countries, while the largest negative gaps appear in countries where women's employment remains more concentrated in agriculture and other less-exposed occupations.
Among the 141 countries in the main exposure panel, 138 have reliable sex-disaggregated occupational comparisons.
Egypt, Micronesia, and Wallis and Futuna remain in the national exposure panel, but are excluded from this gender-gap analysis because their sex-disaggregated occupational cells do not meet the reliability criteria used for the appendix figure.

\begin{figure}[ht]
    \centering
    \includegraphics[width=0.86\linewidth]{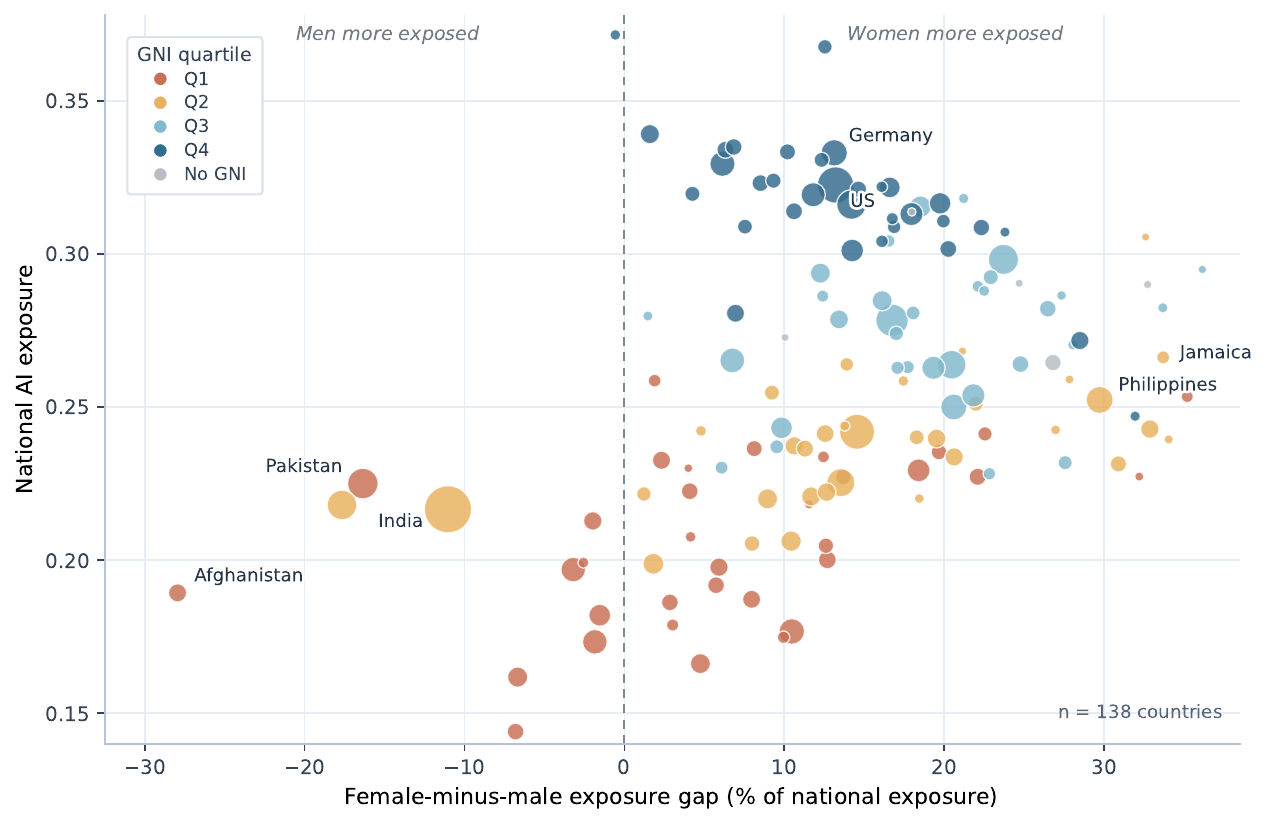}
    \caption{\textbf{Gender differences in national AI exposure.}
    Positive values indicate that female exposure exceeds male exposure, measured as $({n}_{female} - {n}_{male}) / {n}_{total}$.
    Points are countries with sex-disaggregated occupational employment data; colors indicate GNI per capita quartiles and bubble sizes scale with national employment.}
    \label{fig:appendix_gender_exposure_gap}
\end{figure}

\subsection{Predicting National AI Exposure}
\label{app:predictivity}

Beyond the main white-collar specification, we test whether national exposure can be predicted from broader national and occupational covariates.
Among basic national covariates, white-collar share is the strongest single predictor; log GNI and internet penetration are also positively associated with exposure, but explain less cross-country variation on their own. We construct the cognitive-minus-physical (CMP) score as a task-composition diagnostic from O*NET 30.2.
At the ISCO-08 2-digit occupation level, we map O*NET SOC occupations to ISCO-08 and compare O*NET occupational properties with the Gmyrek et al. exposure scores.

The cognitive side is the average of the five O*NET properties most positively associated with exposure: Working with Computers, Written Comprehension, Reading Comprehension, Administrative, and Computers and Electronics.
The physical side is the average of the five properties most negatively associated with exposure: Trunk Strength, Stamina, Performing General Physical Activities, Static Strength, and Gross Body Coordination.
For each occupation, CMP is the cognitive average minus the physical average; for each country, national CMP is the employment-share-weighted average CMP across its measured ISCO-08 occupations.
Higher CMP therefore means that a country's employment is concentrated in occupations with more computer/document/information-processing content and less physical/manual content. This O*NET-derived CMP score provides a more task-grounded measure of labor composition and predicts exposure more strongly than the coarse white-collar split, indicating that the underlying cognitive and physical organization of work is the main driver rather than the white-collar label itself.

\begin{table}[ht]
\centering
\caption{Predicting national AI exposure.}
\label{tab:appendix_predicting_exposure_regressions}
\scriptsize
\setlength{\tabcolsep}{3.5pt}
\resizebox{\linewidth}{!}{%
\begin{tabular}{lcccccccc}
\toprule
 & (1) & (2) & (3) & (4) & (5) & (6) & (7) & (8) \\
\midrule
Specification & WC & GNI & Internet & CMP & WC+GNI & WC+Internet & WC+CMP & Full \\
\midrule
White-collar share & 0.242*** &  &  &  & 0.196*** & 0.212*** & -0.046*** & -0.038** \\
 & (0.007) &  &  &  & (0.011) & (0.008) & (0.017) & (0.015) \\
log GNI &  & 0.036*** &  &  & 0.009*** &  &  & 0.004* \\
 &  & (0.002) &  &  & (0.002) &  &  & (0.002) \\
Internet penetration &  &  & 0.157*** &  &  & 0.031*** &  & -0.007 \\
 &  &  & (0.010) &  &  & (0.009) &  & (0.007) \\
Cognitive-minus-physical score &  &  &  & 0.105*** &  &  & 0.124*** & 0.117*** \\
 &  &  &  & (0.001) &  &  & (0.007) & (0.007) \\
\midrule
Observations & 141 & 135 & 130 & 134 & 135 & 130 & 134 & 124 \\
$R^2$ & 0.907 & 0.772 & 0.669 & 0.973 & 0.928 & 0.925 & 0.975 & 0.979 \\
\bottomrule
\end{tabular}%
}
\vspace{0.25em}
\begin{minipage}{0.94\linewidth}
\footnotesize
\emph{Note:} Entries report OLS coefficients with HC1 robust standard errors in parentheses. The dependent variable is national AI exposure. Internet penetration is measured as a 0--1 share. Each column uses the complete-case sample for the variables in that specification. *** $p<0.01$, ** $p<0.05$, * $p<0.10$.
\end{minipage}
\end{table}

\subsection{Predicting National AI Adoption}

We also estimate the full set of models that include or exclude white-collar share, log GNI, and national AI exposure for each of the three observed adoption outcomes.
National AI exposure strongly predicts adoption on its own across all three datasets.
However, once white-collar share and log GNI are included, adding national AI exposure contributes essentially no additional explanatory power and the full-model exposure coefficient is not statistically distinguishable from zero.
This suggests that exposure is useful as a structural summary, while the observed cross-country adoption relationship is largely aligned with broader national labor composition and income.

\begin{table}[H]
\centering
\caption{Predicting national AI adoption: full regression specifications.}
\label{tab:appendix_predicting_adoption_regressions}
\label{tab:appendix_predicting_adoption_regressions_openai}
\label{tab:appendix_predicting_adoption_regressions_microsoft}
\scriptsize
\setlength{\tabcolsep}{1.7pt}
\renewcommand{\arraystretch}{0.84}
\begin{tabular*}{\linewidth}{@{\extracolsep{\fill}}lcccccccc@{}}
\toprule
 & (1) & (2) & (3) & (4) & (5) & (6) & (7) & (8) \\
\midrule
\multicolumn{9}{@{}l}{\textit{Panel A: Anthropic Claude} (dependent variable: log10(Claude usage / 100k WAP))} \\
Specification & \shortstack{Intercept\\only} & \shortstack{White-collar\\only} & \shortstack{log GNI\\only} & \shortstack{Exposure\\only} & \shortstack{White-collar\\+ log GNI} & \shortstack{White-collar\\+ exposure} & \shortstack{log GNI\\+ exposure} & \shortstack{Full\\model} \\
White-collar share &  & 2.761*** &  &  & 1.123*** & 1.878*** &  & 1.112*** \\
 &  & (0.123) &  &  & (0.270) & (0.545) &  & (0.423) \\
log GNI &  &  & 0.471*** &  & 0.305*** &  & 0.336*** & 0.304*** \\
 &  &  & (0.020) &  & (0.051) &  & (0.061) & (0.059) \\
National AI exposure &  &  &  & 10.643*** &  & 3.602* & 3.561*** & 0.056 \\
 &  &  &  & (0.453) &  & (2.142) & (1.296) & (1.986) \\
Observations & 114 & 114 & 114 & 114 & 114 & 114 & 114 & 114 \\
$R^2$ & 0.000 & 0.792 & 0.837 & 0.769 & 0.863 & 0.799 & 0.854 & 0.863 \\
\addlinespace[0.35em]
\midrule
\multicolumn{9}{@{}l}{\textit{Panel B: OpenAI Signals} (dependent variable: OpenAI rank percentile)} \\
Specification & \shortstack{Intercept\\only} & \shortstack{White-collar\\only} & \shortstack{log GNI\\only} & \shortstack{Exposure\\only} & \shortstack{White-collar\\+ log GNI} & \shortstack{White-collar\\+ exposure} & \shortstack{log GNI\\+ exposure} & \shortstack{Full\\model} \\
White-collar share &  & 1.343*** &  &  & 0.571*** & 0.834*** &  & 0.686*** \\
 &  & (0.056) &  &  & (0.105) & (0.212) &  & (0.144) \\
log GNI &  &  & 0.215*** &  & 0.137*** &  & 0.152*** & 0.144*** \\
 &  &  & (0.009) &  & (0.017) &  & (0.025) & (0.021) \\
National AI exposure &  &  &  & 5.215*** &  & 2.084*** & 1.761*** & -0.634 \\
 &  &  &  & (0.219) &  & (0.777) & (0.590) & (0.724) \\
Observations & 88 & 88 & 88 & 88 & 88 & 88 & 88 & 88 \\
$R^2$ & 0.000 & 0.828 & 0.874 & 0.812 & 0.908 & 0.839 & 0.891 & 0.909 \\
\addlinespace[0.35em]
\midrule
\multicolumn{9}{@{}l}{\textit{Panel C: Microsoft AI Diffusion} (dependent variable: MS GenAI adoption Q1 2026 (\% WAP))} \\
Specification & \shortstack{Intercept\\only} & \shortstack{White-collar\\only} & \shortstack{log GNI\\only} & \shortstack{Exposure\\only} & \shortstack{White-collar\\+ log GNI} & \shortstack{White-collar\\+ exposure} & \shortstack{log GNI\\+ exposure} & \shortstack{Full\\model} \\
White-collar share &  & 53.981*** &  &  & 37.123*** & 45.650*** &  & 39.374*** \\
 &  & (4.045) &  &  & (8.541) & (11.606) &  & (11.082) \\
log GNI &  &  & 8.041*** &  & 2.990** &  & 3.908** & 3.094* \\
 &  &  & (0.752) &  & (1.381) &  & (1.627) & (1.639) \\
National AI exposure &  &  &  & 203.888*** &  & 34.030 & 116.761*** & -11.584 \\
 &  &  &  & (16.654) &  & (38.948) & (38.042) & (50.533) \\
Observations & 106 & 106 & 106 & 106 & 106 & 106 & 106 & 106 \\
$R^2$ & 0.000 & 0.651 & 0.599 & 0.611 & 0.670 & 0.653 & 0.641 & 0.671 \\
\bottomrule
\end{tabular*}
\vspace{0.25em}
\begin{minipage}{0.96\linewidth}
\footnotesize
\emph{Note:} Entries report OLS coefficients with HC1 robust standard errors in parentheses. The table contains 24 models: three adoption outcomes times eight predictor specifications. Each panel uses the outcome's complete-case country sample across white-collar share, log GNI, and national AI exposure. *** $p<0.01$, ** $p<0.05$, * $p<0.10$.
\end{minipage}
\end{table}



\end{document}